\documentclass[aps,nofootinbib,longbibliography,notitlepage,showkeys]{revtex4-1}

\usepackage[totalheight = 23cm, totalwidth = 17cm]{geometry}

\usepackage{graphicx}
\usepackage{amsmath, amssymb, graphics, epsfig, graphicx}
\usepackage{dcolumn}
\usepackage{bm}
\usepackage{hyperref}
\usepackage{epstopdf}

\usepackage{color}
\usepackage[section]{placeins}

\newcommand{\nc}{\newcommand}
\nc{\ba}{\begin{eqnarray}}
\nc{\ea}{\end{eqnarray}}

\nc{\be}{\begin{eqnarray}}
\nc{\ee}{\end{eqnarray}}

\nc{\bfk}{{\bf k }}
\nc{\bfx}{{\bf x }}
\nc{\pfp}{{\bf{p}}}
\nc{\bfq}{{\bf{q}}}
\nc{\tbf}{\textbf}
  
\nc{\calP}{  { \cal P} }  
\nc{\calR}{  { \cal R} }  
\nc{\im}{ \mathrm{Im} }
\nc{\sg}{ \mathrm{sgn} }

\nc{\com}[1]{ \textcolor{blue}{(#1)}}

\begin{document}

\title{Inflationary power asymmetry from primordial domain walls}

\author{Sadra Jazayeri}
\email{sadraj@ipm.ir}
\affiliation{Department of Physics, Sharif University of Technology, Tehran, Iran }
\author{Yashar Akrami}
\email{yashar.akrami@astro.uio.no}
\affiliation{Institute of Theoretical Astrophysics, University of Oslo\\
       P.O. Box 1029 Blindern, N-0315 Oslo, Norway}
\author{Hassan Firouzjahi}
\email{firouz@ipm.ir}
\affiliation{School of Astronomy, Institute for Research in Fundamental Sciences (IPM) \\ P.~O.~Box 19395-5531, Tehran, Iran}
\author{Adam R. Solomon}
\email{a.r.solomon@damtp.cam.ac.uk}
\author{Yi Wang}
\email{y.wang@damtp.cam.ac.uk}
\affiliation{DAMTP, Centre for Mathematical Sciences, University of Cambridge\\
       Wilberforce Rd., Cambridge CB3 0WA, UK}

\begin{abstract}

We study the asymmetric primordial fluctuations in a model of inflation in which translational invariance is broken by a domain wall. We calculate the corrections to the power spectrum of curvature perturbations; they are anisotropic and contain dipole, quadrupole, and higher multipoles with non-trivial scale-dependent amplitudes. Inspired by observations of these multipole asymmetries in terms of two-point correlations and variance in real space, we demonstrate that this model can explain the observed anomalous power asymmetry of the cosmic microwave background (CMB) sky, including its characteristic feature that the dipole dominates over higher multipoles. We test the viability of the model and place approximate constraints on its parameters by using observational values of dipole, quadrupole, and octopole amplitudes of the asymmetry measured by a local-variance estimator. We find that a configuration of the model in which the CMB sphere does not intersect the domain wall during inflation provides a good fit to the data. We further derive analytic expressions for the corrections to the CMB temperature covariance matrix, or angular power spectra, which can be used in future statistical analysis of the model in spherical harmonic space.

\end{abstract}


\maketitle

\tableofcontents

\section{Introduction}

Cosmic inflation is the leading paradigm for describing the very early Universe. The simplest models of inflation, based on a scalar field slowly rolling down a flat potential, predict
nearly scale-invariant, adiabatic, and Gaussian perturbations. These predictions agree exquisitely with precision observations of the cosmic microwave background (CMB) 
\cite{Ade:2013zuv, Ade:2013uln}.

Despite the remarkable consistency of the simplest inflationary models with cosmological observations, there
are curious large-scale (i.e., low-multipole) anomalies in the observed CMB maps (see, e.g., Refs. \citep{Tegmark:2003ve,deOliveiraCosta:2003pu,Vielva:2003et,Larson:2004vm,Land:2004bs,Land:2005ad,McEwen:2004sv,McEwen:2006yc,Jaffe:2006sq,Hinshaw:2006ia,Spergel:2006hy,Cruz:2006fy,Bridges:2006mt,Copi:2006tu,Land:2006bn,Bernui:2005pz,Bernui:2008ve,Pietrobon:2008rf}), which, if future observations confirm them and suggest that they have a primordial origin, would require a non-trivial model of inflation. We will focus in particular on the growing evidence for the existence of a {\it power asymmetry} in 
the CMB sky as observed by the {\it Wilkinson Microwave Anisotropy Probe} ({\it WMAP}) \cite{Bennett:2003bz} and {\it Planck} \cite{Ade:2013ktc} experiments \cite{Eriksen:2003db,Hansen:2004vq,Park:2003qd,Eriksen:2007pc,hansen2009,hoftuft2009,Axelsson:2013mva,Ade:2013nlj,Akrami:2014eta, Notari:2013iva}.

A widely-used phenomenological model for the observed power asymmetry is a dipole modulation \cite{Gordon:2006ag},
\ba
\frac{\Delta T}{T}|_{\textrm{mod}}(\hat{\bf{n}})=(1+A \hat{\bf{n}}\cdot\hat{\bf{p}})\frac{\Delta T}{T}|_{\textrm{iso}}(\hat{\bf{n}})\label{eq:dipole},
\ea
where $\frac{\Delta T}{T}|_{\textrm{iso}}(\hat{\bf{n}})$ and $\frac{\Delta T}{T}|_{\textrm{mod}}(\hat{\bf{n}})$ are the isotropic and observed (modulated) temperature fluctuations in a direction $\hat{\bf{n}}$ on the sky, respectively, $A$ is the dimensionless amplitude of the modulation, and $\hat{\bf{p}}$ is a preferred direction. A typical value for $A$ measured by {\it WMAP} and {\it Planck} is $A\sim0.07$ \cite{hoftuft2009,Ade:2013nlj}. However, this model does not seem to be applicable on all scales: dipole asymmetry has been observed only at low multipoles ($\ell \lesssim 64$) and seems to vanish on small scales ($\ell \gtrsim 600$) \cite{Ade:2013nlj,Flender:2013jja,Akrami:2014eta}, as is confirmed by observations of the large-scale structure of the Universe \cite{Hirata:2009ar}. This indicates that the asymmetry cannot be described exactly by Eq.~(\ref{eq:dipole}), i.e., by an all-sky dipole modulation with a scale-independent amplitude (see, however, Ref.~\cite{Quartin:2014yaa}). This dependence on scale should be taken into account in any theoretical attempts to model the power asymmetry.  

The idea of generating power asymmetry from a long-mode modulation \cite{Erickcek:2008sm,Erickcek:2008jp,Dai:2013kfa}
has been studied extensively in the literature \cite{Donoghue:2007ze,Lyth:2013vha,Wang:2013lda,Liu:2013kea,McDonald:2013aca, Liddle:2013czu,Mazumdar:2013yta, Cai:2013gma,Chang:2013lxa,Kohri:2013kqa,McDonald:2013qca,  Kanno:2013ohv,Liu:2013iha,Chang:2013mya,McDonald:2014lea}. In this scenario, a superhorizon mode modulates the background inflationary
parameters and introduces a dipolar asymmetry. However, it is difficult to generate a large enough asymmetry in simple single-field models of inflation. This is because the amplitude of the dipole asymmetry is controlled by the amount of non-Gaussianity in the squeezed limit \cite{Namjoo:2013fka, Abolhasani:2013vaa, Namjoo:2014nra}. 
In single-field models of inflation which satisfy Maldacena's consistency condition \cite{Maldacena:2002vr},
the non-Gaussianity parameter $f_\mathrm{NL}$ is too small to generate a sufficiently large amplitude for the dipole asymmetry. As a result, in order to explain the observed power asymmetry, one needs to violate Maldacena's consistency condition by going beyond single-field models of inflation. This can be achieved either by considering a multiple-field model of inflation or by assuming non-vacuum initial conditions \cite{Firouzjahi:2014mwa}.  In addition,
it has not been easy to generate a {\it scale-dependent} dipole asymmetry in these models \cite{Erickcek:2009at}.
Therefore, it is  still an open question how to generate a scale-dependent dipole-like asymmetry with the desired amplitude from the simplest well-motivated inflationary models. 

In this work, we consider a model of inflation in which translational invariance is broken by the presence of a domain wall. A domain wall can arise before or during inflation in many ways, such as the Kibble mechanism \cite{Kibble:1976sj} in models with a discrete $\mathbb Z_2$ symmetry breaking, bubble nucleation \cite{Guth:1980zm}, or bifurcation during inflation \cite{Li:2009sp,Wang:2013vxa}. We calculate the corrections to the power spectrum of primordial fluctuations in the presence of a domain wall, as well as to the two-point correlation function and the variance in real space. The corrections turn out to have a non-trivial scale-dependence and in general can generate dipolar as well as higher-order modulations to the CMB temperature fluctuations. We study these predictions and compare them to the observed properties of the asymmetry provided by existing CMB experiments. The data to which we compare our predictions include, in particular, the amplitudes of the dipole and higher multipoles for the local-variance maps presented in Ref.~\cite{Akrami:2014eta} for both {\it WMAP} and {\it Planck} temperature data. Our results show that the presence of a domain wall is able to successfully explain the asymmetry in the data when we choose appropriate (and reasonable) values for the model's two free parameters: the tension of the wall and its distance from the CMB sphere. Although the anisotropy structure of the model is quite sophisticated, it is able to provide a dominant dipolar asymmetry with an effectively scale-dependent amplitude that is consistent with observations. We further calculate the full CMB covariance matrix, and discuss the prospects for an extensive statistical analysis of the model based on the CMB angular power spectra in spherical harmonic space. We note that a similar question has 
been studied in Ref.~\cite{Carroll:2008br} in which the effects of violation of translational invariance during inflation in the presence of a preferred point, line or plane have been investigated. 

This article is organized as follows. In section \ref{model} we present our model and its configuration, the background metric during inflation in the presence of a domain wall, and the relevant background expressions and equations. We then present the curvature perturbations in section \ref{power} and derive the corrections to the primordial scalar power spectra induced by the domain wall. In section \ref{corr-general} we derive predictions for the two-point correlations and the variance in real space, focusing in 
subsection \ref{corr-theory} on theoretical calculations, while discussing the implications of this model for observations in subsection \ref{corr-obs}. We compare our predictions for the multipole structure of the variance to CMB observations, finding results that are consistent with observations while providing tight constraints on the parameters of the model. We derive the CMB angular power spectra in section \ref{ang-power}, where we discuss how they can be used in the future for a proper statistical analysis of the model using CMB data. We conclude in section \ref{conc}.

\section{Background geometry in the presence of a domain wall}
\label{model}

In this section we present and discuss our background setup. We would like to study the effects of a domain wall present during inflation. This breaks translational invariance and leads to a modification of the curvature spectrum.

We assume inflation is driven by a single scalar inflaton $\phi$ slowly rolling down a flat potential $V(\phi)$.  The domain wall has surface energy density $\sigma$. We assume  the dominant source of energy is from the inflaton potential, so the energy of the wall over a Hubble radius is small compared to that of the inflaton potential, $\sigma \ll V/H\sim\sqrt{V}M_\mathrm{Pl}$, where $M_\mathrm{Pl}$ is the reduced Planck mass.  
This allows us to treat the contribution of the domain wall perturbatively. We assume that the domain wall is created dynamically during inflation and subsequently disappears either during or at the end of inflation. At this level, we do not provide a dynamical mechanism for the formation and annihilation of the domain wall, which is an interesting question but beyond the scope of the present work.

The wall is assumed to be extended in the $x-y$ plane. Therefore, translational symmetry along the direction perpendicular to the domain wall is broken, and we are left with a two-dimensional 
symmetry in the directions parallel to it. This indicates a violation of the Copernican principle
and can provide a setup for the mechanism considered in Ref.~\cite{Kamionkowski:2014faa}  to generate dipole asymmetry in the 
CMB temperature map from tensor polarizations. 

To simplify the analysis, we assume the inflationary background is a de Sitter space, generated by a constant potential $V$. This is of course a simplification: one can consider 
the more realistic case in which $V$ has a mild dependence on $\phi$, but this brings slow-roll suppressed corrections. Since the anisotropies which interest us are generated even in the presence of a constant potential, we neglect these slow-roll corrections. Furthermore, because the domain wall's energy is subdominant to the inflaton's, we can model its gravitational effect as a small perturbation to de Sitter space. Therefore, our first job is to determine the spacetime metric in the presence of a domain wall in a de Sitter background.  This analysis was performed in Refs. \cite{Wang:2010mb, Wang:2011pb}. Here we review the main results.

\subsection{The background metric}
\label{back-metric}

Assuming the domain wall is extended in the $x-y$ plane, we consider the ansatz
\ba
ds^2 = \frac{1}{f( \tau,z)^2} \left( - d \tau^2 + d \bfx^2 \right),
\ea
where $\tau$ is conformal time. With a constant scalar potential and the domain wall localized at $z=0$, the   energy-momentum  tensor is 
\ba
\label{T-munu}
T^{\mu}{}_{\nu}=- V \delta^{\mu}{}_{\nu}- \frac{\sigma}{\sqrt{g_{zz}}} \mathrm{diag}\left(1,1,1,0 \right)  \delta(z).
\ea
The factor $1/\sqrt{g_{zz}}$ comes from the determinant of the metric on the worldvolume
of the wall, which is three dimensional while the bulk space is four dimensional. The notation $ \mathrm{diag} \left(1,1,1,0 \right) $ means  that the wall is extended along the $\tau$, $x$, and $y$ directions. 

The nonzero components of the Einstein tensor $G^\mu{}_\nu$  are given by
\ba
\label{00}
G^0{}_0 &=& -3 \frac{\partial^2 f}{\partial \tau^2} +  3 \left( \frac{\partial f}{\partial z}\right)^2 - 2 f \frac{\partial^2 f}{\partial z^2}, \\
\label{0-1}
G^1{}_1 &=& 2 f  \frac{\partial^2 f}{\partial \tau^2} + G^0{}_0, \\
\label{0-3}
G^3{}_3 &=& 2 f \left(\frac{\partial^2 f}{\partial \tau^2} +  \frac{\partial^2 f}{\partial z^2} \right) + G^0{}_0, \\
\label{03}
G^0{}_3 &=& -2 f \frac{\partial^2 f}{\partial \tau \partial z}.
\ea
We would like to solve the Einstein equation, $G^\mu{}_\nu = T^\mu{}_\nu/M_\mathrm{Pl}^2$.  From the symmetry of the system we have $G^1{}_\mu = G^2{}_\mu$. 
In the bulk, defined by $z \neq 0$, we have $T^\mu{}_\nu = -V \delta^\mu{}_\nu$, implying 
$G^0{}_0=G^1{}_1= G^3{}_3$. Therefore, from Eq.~(\ref{0-1}) we get $ \frac{\partial^2f}{\partial \tau^2}=0$ while from Eq.~(\ref{0-3})
we have $  \frac{\partial^2 f}{\partial \tau^2} +   \frac{\partial^2 f}{\partial z^2} =0$. Finally, from
Eq.~(\ref{03}) we get $\frac{\partial^2 f}{\partial \tau \partial z}=0$.  These equations imply that $f(\tau, z)$ is  linear in $\tau$ and $z$.  Since we assume the background is symmetric under reflection with respect to the plane $z=0 $ we have 
\ba
f= \alpha_1 \tau + \alpha_2 | z|,
\ea
where $\alpha_1$ and $\alpha_2$ are two constants of integration to be determined. Note that, as in an exact de Sitter solution, we assume $\tau <0$. 

We have derived the metric up to two constants of integration, $\alpha_{1,2}$. Let us now determine these constants. From Eq.  (\ref{00}) we have 
\ba
\label{00-eq}
3 \alpha_1^2 - 3 \alpha_2^2 = \frac{V}{M_\mathrm{Pl}^2}  .
\ea 
To fix $\alpha_1$ and $\alpha_2$ we have to impose boundary conditions at the position of the
domain wall, $z=0$.  Considering the singular parts of Eq.  (\ref{00})  we have
\ba
- 2 f  \frac{\partial^2 f}{\partial z^2} \sim -\frac{\sigma}{\sqrt{g_{zz}}} \delta (z) = -| f |\sigma \delta(z).
\ea
Performing the matching condition, and noting that $\tau <0$, we find
\ba
\label{alpha2}
\alpha_2 = \frac{- \sigma}{4 M_\mathrm{Pl}^2} \sg(\alpha_1).
\ea
The domain wall tension should have $\sigma >0$ to be physical. As a result, we see that  $\alpha_2$ has the opposite sign to  $\alpha_1$. 
Plugging this into Eq.~(\ref{00-eq}) we obtain
\ba
\alpha_1 = \pm \sqrt{   \frac{V }{3 M_\mathrm{Pl}^2} +  \frac{ \sigma^2}{16 M_\mathrm{Pl}^4}}.
\ea

In conclusion, the de Sitter metric perturbed by a low-energy domain wall is
\ba
\label{metric2}
ds^2 = \frac{1}{\alpha^2 \left(  \tau - \beta |z|  \right)^2}  \left(- d\tau^2 + d \bfx^2 \right),
\ea
where
\ba
\alpha^2 \equiv \frac{V }{3 M_\mathrm{Pl}^2} +  \frac{ \sigma^2}{16 M_\mathrm{Pl}^4} , \qquad
\beta \equiv  \frac{\sigma}{4 M_\mathrm{Pl}^2|\alpha|}.
\ea
Notice that, because of the positivity of the domain wall's tension, $\beta$ is always positive. This originates from Eq.~(\ref{alpha2}), which constrains $\alpha_1$ and $\alpha_2$ to have opposite signs. We have already assumed $\sigma \ll \sqrt{V}M_\mathrm{Pl}$ in order to ensure the domain wall's energy is subdominant on Hubble scales, so we can write $\beta$ approximately as
\ba
\beta \approx \frac{\sqrt3\sigma}{4M_\mathrm{Pl}V^{1/2}} \approx \frac{\sigma}{4HM_\mathrm{Pl}^2}, \label{eq:betaval}
\ea
where $H \approx \sqrt{V/3M_\mathrm{Pl}^2}$ is the Hubble rate in the slow-roll limit. Therefore we can interpret $\beta$ as the energy density of the domain wall in terms of the other relevant energy scales; it is dimensionless and is a small parameter.

\subsection{Boundary and conformal time}

The space-time with metric (\ref{metric2}) contains a boundary at 
\ba
\label{hor-1}
\tau-\beta |z|=0.
\ea
This reduces to the usual de Sitter future boundary $\tau =0$ in the case where $\beta=0$. One can check that the  proper time of a geodesic observer reaches infinity as the observer approaches the boundary $\tau = \beta|z|$. 
In this view, the observable universe can be interpreted as $\tau-\beta |z|<0$, as the space beyond 
this region is not accessible to a geodesic observer.
 
If one changes the coordinates on one side of the domain wall (for instance $z>0$) as \cite{Wang:2011pb} 
\ba
\tilde{z}\equiv\frac{z-\beta \tau}{\sqrt{1-\beta^2}}, \\ \nonumber 
\tilde{\tau}\equiv\frac{\tau-\beta z}{\sqrt{1-\beta^2}} ,
\ea
the metric becomes exact de Sitter on that side, and all geodesics approach a constant $\tilde{z}$. An observer's proper time asymptotes to infinity as they approach the boundary. From this point of view it is clear that this boundary represents the end of our accessible universe.

The time coordinate $\tau$ in the metric (\ref{metric2}) is not, however, a good conformal time. First, the boundary given in Eq.~(\ref{hor-1}) depends on $z$, i.e., the boundary cannot be moved to a constant-time surface. This should be compared to the boundary in pure de Sitter space which is everywhere given by $\tau=0$. The second problem with the coordinate system in (\ref{metric2}) is that  the metric cannot be expanded perturbatively with respect to $\beta$.
Recall from Eq.~(\ref{eq:betaval}) that $\beta$ measures the smallness of the domain wall energy in a Hubble volume, compared to the energy of the inflaton, so by construction it is small. The effects of the domain wall, in the metric and in physical observables such as the curvature perturbations, should therefore be perturbative in $\beta$. This is not transparent from the metric (\ref{metric2}). 

\begin{figure}
\hspace{2cm}
\includegraphics[ width=0.7\linewidth]{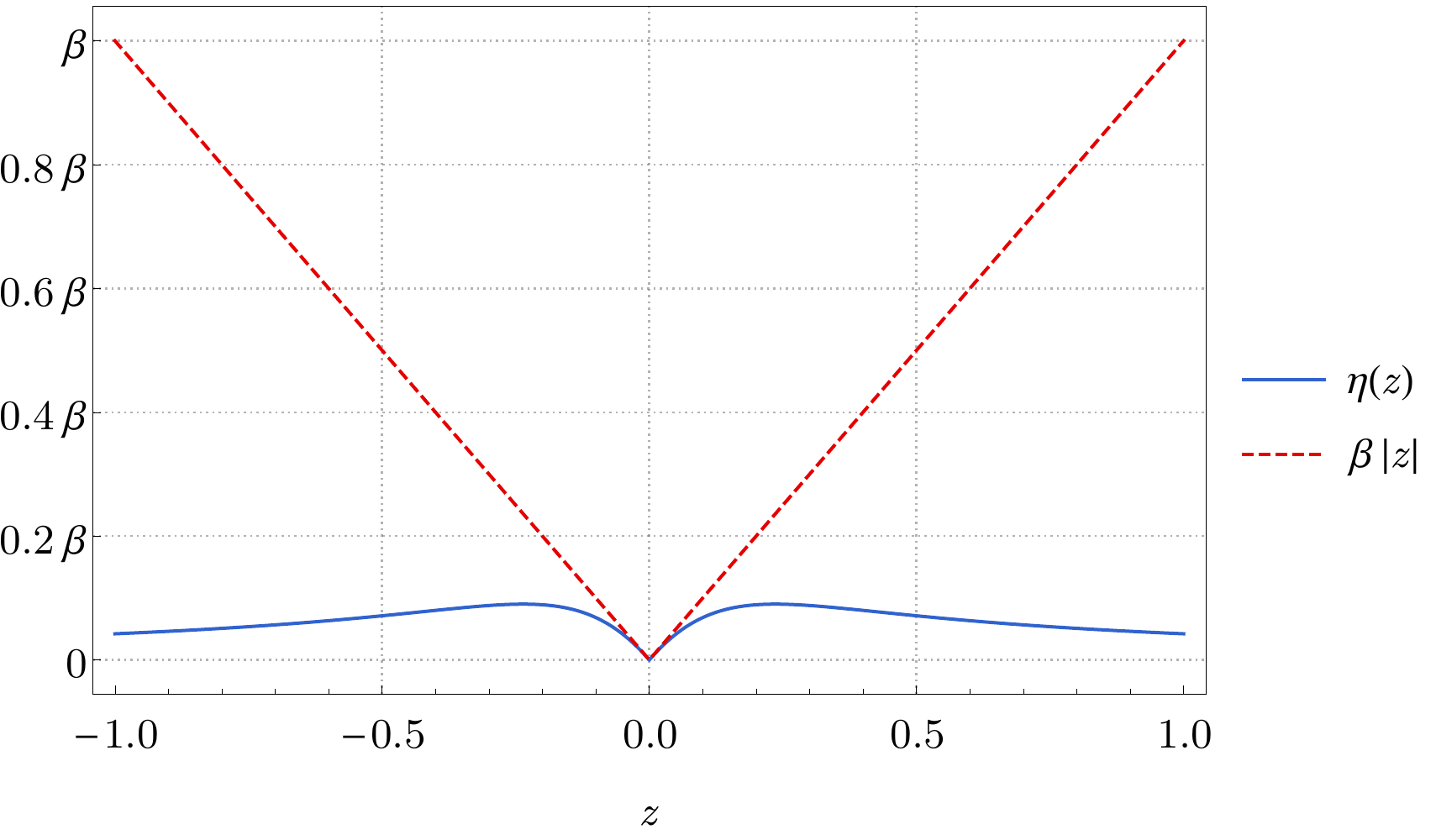}
\caption{ 
Boundary as defined by Eq.~(\ref{hor-2}) in the $\eta-z$ coordinates (solid curve).  By making $\delta$ arbitrarily small, the boundary curve approaches the axis $\eta=0$. The dashed lines represent $\eta = \beta | z|$. For concreteness we have plotted the curve with $\delta=0.3$.
 }
\label{stop}
\end{figure}

The boundary given in Eq.~(\ref{hor-1}) indicates that one has to use a new conformal time $\eta$ such that the boundary (\ref{hor-1})  is given by $\eta=0$. However, the function $| z|$ makes 
the transformation $\tau-\beta |z| \rightarrow \eta$ problematic as this coordinate transformation is not differentiable at $z=0$. In order to get rid of this difficulty, we define a new conformal time
\ba
\label{eta-tau-new}
{\eta}\equiv\tau -\frac{\beta z^2}{\sqrt{z^2+\delta^2}},
\ea
and leave the spatial coordinates unchanged. Here we have introduced a small parameter  $\delta \rightarrow 0$ to approximate the function $| z|$ by the smooth function $\frac{ z^2}{\sqrt{z^2+\delta^2}}$. By taking $\delta $ arbitrarily close to zero this approximation becomes more and more precise. With this coordinate transformation, the boundary in Eq.~(\ref{hor-1}) is mapped to
\ba
\label{hor-2}
{\eta}= \beta |{z}| - \frac{ \beta z^2}{\sqrt{z^2+\delta^2}}.
\ea
For $ |z| \gg \delta $, the boundary is simply mapped to $\eta \simeq 0$. Near the origin 
$| z| \leq \delta$ the boundary deviates from the  axis $\eta =0$ with a height which is proportional to $\delta$. By taking $\delta$ arbitrarily close to zero, the boundary given in Eq.~(\ref{hor-2}) approaches the surface of constant time, $\eta =0$, more and more accurately for all values of $z$. In this limit, $\eta$ satisfies our second criterion for being a good conformal time coordinate, as the boundary is determined by the constant hypersurface $\eta=0$. A schematic view of the boundary defined in Eq.~(\ref{hor-2}) is presented in Fig.~\ref{stop}.

In terms of this new conformal time, the metric (\ref{metric2}) becomes
\ba
\label{metric3}
ds^2=\frac{1}{\alpha^2 \left(\eta+\frac{\beta z^2}{\sqrt{z^2+\delta^2}}-\beta |z| \right)^2} \left[ -\left(d \eta + \frac{2\beta z dz}{\sqrt{z^2+\delta^2}}-\frac{\beta z^3 dz}{(z^2+\delta^2)^{\frac{3}{2}}} \right)^2+
d\textbf{x}^2\right].
\ea
One can check that this metric is an exact solution of the Einstein equation with the 
energy-momentum tensor  given by Eq.~(\ref{T-munu}), as it should be. The metric
(\ref{metric3}) also satisfies our first criterion for a good conformal time: it can be perturbed with respect to $\beta$ for all values of $z$. This is because the denominator is just $\eta^{-2}$ in the small $\delta$ limit (with an error of ${\cal O}(\delta^2)$, which can be made arbitrarily small), leaving us with only an off-diagonal term $\sim\beta d\eta dz$, which can be treated as a small perturbation to exact de Sitter.

To simplify matters, let us consider the limit $\delta \rightarrow 0$ in some more depth. In this limit we have
\ba
\frac{\beta z^2}{\sqrt{z^2+\delta^2}} -\beta |z| = \frac{- \beta \delta^2}{2 | z|} +  {\cal O}(\delta^4) \quad \quad (\delta \rightarrow 0).
\ea
Therefore, for $| \eta | > \beta \delta^2/2 |z| $ one can safely approximate the denominator of the metric (\ref{metric3}) by $\eta^{-2}$ as in an exact de Sitter background. In particular, denoting the time at the end of inflation by $\eta_e$, this approximation is valid for $| z| > \delta^2/2 | \eta_e|$. Taking $\delta$ arbitrarily close to zero, one can cover the entire range of $z$ with the exception of $z=0$, the position of the domain wall.  

Finally, in the limit $\delta\to0$, the metric (\ref{metric3}) simplifies to 
\ba
\label{metric4}
ds^2=\frac{1}{\alpha^2 \eta^2}\left(-d\eta^2 - 2\beta  \sg(z)d\eta dz+(1-\beta^2)dz^2+dx^2+dy^2 \right).
\ea
The leading effect of the new time coordinate is to induce the off-diagonal term $\sim\beta d\eta dz$. We note that unlike the metric (\ref{metric3}), which is an exact solution of the Einstein equations in the presence of the domain wall, the metric (\ref{metric4}) cannot be an exact solution. This is because the latter is obtained by setting $\delta =0$, which from Eq. \ref{eta-tau-new}, corresponds to the coordinate transformation $\eta = \tau -\beta |z|$ which is not regular at $z=0$. We have specifically checked that the metric (\ref{metric4}) induces errors of ${\cal O}(\beta^3)$ and higher  orders in solving the Einstein equations  with the matching conditions. 
However, as we shall see, the corrections to the power spectrum induced by the domain wall will be at the order of $\beta$, and for small values of $\beta$ the higher-order errors induced by the metric  (\ref{metric4}) can be neglected. 

Equipped with the metric (\ref{metric4}) we are ready to treat the system perturbatively in terms of $\beta$. In particular, we neglect terms of order $\beta^2$ in $dz^2$ and in the rest of our analysis. As in conventional models of inflation we work in the regime $ -\infty<\eta <\eta_e$, and take the end of inflation to be approximately the de Sitter boundary, $\eta_e \rightarrow 0$. 

\section{Anisotropies in curvature perturbations} 
\label{power}

In the previous section we calculated the background metric of an inflationary spacetime perturbed by a domain wall, Eq.~(\ref{metric4}). We are now ready to calculate the corrections to the curvature perturbation power spectrum $P_\calR$ induced by the wall. We assume inflation is driven by the inflaton field $\phi$ slowly rolling over the flat potential $V(\phi)=V$. The comoving curvature perturbation associated with the scalar field fluctuations is given by $\calR = H \delta \phi/\dot \phi$, where $H$ is the Hubble rate and $\delta \phi$ represents the inflaton's quantum fluctuations.  To leading order, the correlation function of the curvature perturbation $\calR$ in Fourier space is given by 
\ba
\langle \calR_\bfk \calR_\bfq \rangle = \left( \frac{H}{\dot \phi}  \right)^2 \langle \delta \phi_\bfk \delta \phi_\bfq \rangle, \label{eq:corrfunc}
\ea
where $H$ and $\dot \phi$ are calculated in the absence of both the domain wall and the scalar perturbations, i.e., in the exact de Sitter limit, and $\bfk$ and $\bfq$ are wavevectors. In the following, we refer to this quantity as the power spectrum of curvature perturbations, $P_\calR$,
although, strictly speaking, the power spectrum is the correlation function for $\bfk=\bfq$.

The effects of the domain wall in Eq.~(\ref{eq:corrfunc}) are contained entirely in the quantity $\langle \delta \phi_\bfk \delta \phi_\bfq \rangle$.\footnote{The presence of the domain wall will also introduce corrections to the background quantities $H$ and $\dot \phi$ through the metric (\ref{metric4}). However, we have checked that these corrections are ${\cal O}(\beta \sqrt\epsilon)$, where $\epsilon$ is the slow-roll parameter, so these corrections can be neglected compared to the $\cal O(\beta)$ terms which the domain wall induces in $\langle \delta \phi_\bfk \delta \phi_\bfq \rangle$.} Here the effect of the wall is to modify the background space from exact de Sitter to
the space given by the metric (\ref{metric4}), breaking translational invariance in the $z$ direction. The quantum fluctuations of the inflaton are, as usual, taken to be the fluctuations of massless scalar fields, calculated in the 
background metric (\ref{metric4}). This will produce corrections from the wall in $P_\calR$ which enter at ${\cal O}(\beta )$.

The second-order Lagrangian for the scalar field fluctuations is
\ba
\mathcal L = \sqrt{-g}\left(-\frac{1}{2}g^{\mu\nu} \partial_{\mu} \delta \phi \partial_{\nu} \delta \phi \right),
\ea
where we should use the metric (\ref{metric4}). To $\mathcal{O}(\beta)$, the determinant and inverse of $g_{\mu\nu}$ are given by
\ba
\sqrt{-g}=\frac{1}{ \alpha^4 \eta^4}  , \qquad -g^{00} = g^{11} = g^{22} = g^{33} = \alpha^2\eta^2, \qquad
g^{03}= g^{30}= - \beta \alpha^2 \eta^2  \sg(z).
\ea
Calculating the quadratic Lagrangian density, and noticing that $\alpha^2 \approx V/3M_\mathrm{Pl}^2 = H^2$ to leading order in $\beta$,  we get 
\begin{equation}
\mathcal{L}=\frac{1}{2 H^2 \eta^2 }\left(\delta \phi'^2 -(\nabla \delta \phi)^2 \right)+\frac{ \beta }{ H^2 \eta^2} \sg(z) \delta \phi'\frac{\partial \delta \phi}{\partial z},
\end{equation}
where $'$ denotes a derivative with respect to conformal time, $\eta$. 

To use the perturbative in-in formalism \cite{Weinberg:2005vy,Chen:2009zp,Chen:2010xka,Wang:2013zva} we have to calculate the interaction Hamiltonian. For this we need to calculate the conjugate momentum, given by  
\ba
\Pi=\frac{\delta \phi'}{\eta^2 H^2}+\frac{\beta}{H^2 \eta^2} \sg(z)  \frac{\partial \delta \phi}{\partial z}  .
\ea
Correspondingly, the quadratic Hamiltonian density is
\ba
\label{H-in}
\mathcal{H}=\mathcal{H}_0-\beta \sg(z)\Pi \partial_z \delta \phi,
\ea
where $\mathcal{H}_0$ is the Hamiltonian density of the free field theory corresponding to $\beta=0$. The second term in Eq.~(\ref{H-in}) represents the interaction Hamiltonian density ${\cal H}_I$ to leading order in $\beta$,
\ba 
\label{HI}
\mathcal{H_I}=-\beta  \Pi  \partial_z \delta \phi  \sg(z) = -\frac{\beta}{H^2 \eta^2} \delta \phi'  \partial_z \delta \phi   \sg(z)  .
\ea 
Correspondingly, the interaction Hamiltonian $H_I$ is given by
\ba
\label{H-in2}
H_I =  -\frac{\beta}{H^2 \eta^2} \int d^3\textbf{x} \, \sg(z) \delta \phi ' \partial_z \delta \phi    .
\ea
We calculate the correction to the power spectrum by taking 
${H_I}$ as the leading interaction Hamiltonian.  The power spectrum is calculated in Fourier space, therefore, we need to calculate $H_I$ also in Fourier space. For this purpose, we need the following representation of the sign function in momentum space:
\ba
\sg(z) = \frac{1}{i \pi} \int_{-\infty}^{\infty} \frac{d p}{p} e^{i p z}.
\ea
In addition, the domain wall breaks the three-dimensional translational invariance to a two-dimensional symmetry. Therefore, it is instructive to decompose the momenta $\bfk$ and $\bfq$ into the tangential parts $\bfq_{||}$ and $\bfk_{||}$ and the vertical parts $k_z $ and $q_z$ as follows:
\ba
\bfk = \bfk_{||}  + k_z \hat {\bf z} \quad , \quad 
\bfq = \bfq_{||}  + q_z \hat {\bf z}.
\ea
Plugging these into the expression for $H_I$ in Eq.~(\ref{H-in2}) we get 
\ba
\label{HI-k}
H_{I } = \frac{2 \beta}{H^2 \eta^2 (2 \pi)^4} \int d^2 \bfq_{||}   d q_z   d k_z  \frac{q_z}{k_z + q_z}
\delta \phi(\bfq) \delta \phi' (-\bfq_{||}, k_z)  .
\ea
Note that we have presented the explicit momentum dependence in $\delta \phi'$, for which $\bfk_{||}= - \bfq_{||}$. 

We are now ready to calculate the modification to the curvature perturbation power spectrum using the standard 
in-in formalism.
As described before, the free theory corresponds to quantum fluctuations $\delta \phi$ in the exact
de Sitter background obtained when $\beta=0$. We define the correction to the power spectrum in Fourier space by
\ba
\langle \calR_\bfk \calR_\bfq \rangle = \left( \frac{H}{\dot \phi}  \right)^2 \left(\big\langle \delta \phi_\bfk \delta \phi_\bfq \big\rangle + \delta\big\langle \delta \phi_\bfk \delta \phi_\bfq \big\rangle\right),
\ea
where the first term is the standard, isotropic power spectrum and the second part is the correction from the domain wall, and
\ba 
\label{delta-power0}
\delta \bigg \langle \delta \phi_\bfk \delta \phi_\bfq \bigg \rangle  \equiv 
+i\int_{- \infty}^{\tau_e} \bigg \langle \big[H_I(\eta), \delta \phi_{\bfk} \delta \phi_{\bfq} \big] \bigg\rangle d\eta  .
\ea

Plugging the form of $H_I$ from Eq.~(\ref{HI-k}) into Eq.~(\ref{delta-power0}) yields
\ba
\label{delta-power1}
\delta  \bigg \langle \delta \phi_\bfk \delta \phi_\bfq \bigg \rangle  =  -\frac{4 \beta}{H^2 (2 \pi)^4} 
\int \frac{d\eta}{\eta^2} \int d^2 \bfq'_{||} d q'_z d k'_z   \frac{q'_z}{k'_z + q'_z} 
\im \bigg[ \bigg \langle \delta \phi_{\bfq'} (\eta) \delta \phi'_{\bfk'}(\eta)  \delta \phi_\bfk(\eta_e) \delta \phi_\bfq(\eta_e)  \bigg \rangle 
\bigg],
\ea
where it is understood that $\bfq'_{||} = - \bfk'_{||}$. Calculating the expectation values using the Wick's theorem, we get
 \ba
 \label{delta-power2}
\delta  \bigg \langle \delta \phi_\bfk \delta \phi_\bfq \bigg \rangle  =  -\frac{4 \beta  q_z}{H^2(k_z + q_z)} (2 \pi)^2
\delta^2 (\bfk_{||} + \bfq_{||} )  \int \frac{d\eta}{\eta^2}   
\im \bigg[ \delta \phi_q(\eta) \delta \phi_k'(\eta) \delta \phi_q^*(\eta_e) \delta \phi_k^*(\eta_e)
\bigg]  + k \leftrightarrow q.
 \ea
 The wave function of the free theory is given by
 \ba
 \delta \phi_k = \frac{H}{\sqrt{2 k^3}} (1- i k \eta) e^{i k \eta}.
 \ea
 Plugging this into the integral in Eq.~(\ref{delta-power2}) and using $\eta_e \simeq 0$, the term containing $\im [ ...] $ in Eq.~(\ref{delta-power2}) becomes proportional to  $\im (I_1 + I_2)$ where the integrals $I_1$ and $I_2$ are defined via
 \ba
 I_1 \equiv \int_{-\infty} ^0 \frac{d \eta }{\eta}  e^{i ( k + q) \eta} \quad , \quad
 I_2 \equiv -i q \int_{-\infty}^0 d \eta  e^{i ( k+ q) \eta}. 
 \ea
Using the contour rotation $\eta = -\infty (1- i \epsilon_0)$ with $\epsilon_0 \rightarrow 0^+$, the UV contribution in $I_2$ is canceled, while the IR contribution from $\eta =0$ in $I_2$ is found to be  real. As a result, 
 $\im (I_2)=0$. On the other hand, for $\im(I_1)$ we have
 \ba
 \im (I_1)  =  \int_{-\infty}^0 \frac{d \eta}{\eta} \sin(k \eta) = \frac{\pi}{2}  .
 \ea
 Plugging these results into Eq.~(\ref{delta-power2}) we get 
 \ba
 \delta  \bigg \langle \delta \phi_\bfk \delta \phi_\bfq \bigg \rangle  =  -\frac{ \beta H^2 }{4 k^3 q^3}
 \frac{k^2 q_z + q^2 k_z}{k_z + q_z} ( 2 \pi)^3 \delta^2 (\bfk_{||} + \bfq_{||}).
 \ea
 Finally, using the relation $\calR = \frac{H}{\dot \phi} \delta \phi$, the total power spectrum, including  the isotropic contribution $\calP_0 \equiv \left( \frac{H^2}{2 \pi \dot \phi} \right)^2$, is given by
\ba
\label{power-final}
\langle  \calR_\bfk \calR_\bfq  \rangle  =  \frac{2 \pi^2}{k^3} \calP_0
\left[ (2 \pi)^3 \delta^3(\bfk + \bfq) - (2 \pi)^3  \frac{\beta }{2 q^3} \frac{k^2 q_z + q^2 k_z}{k_z + q_z} \delta^2 (\bfq_{||} + \bfk_{||})\right].
\ea
Eq.~(\ref{power-final}) is the main result of this section. The first term represents the isotropic contribution in the absence of domain wall while the second term encodes the effects of domain wall.
Note that because of the domain wall, translational invariance along the direction perpendicular to the wall is broken. Also note that the leading-order correction to the power spectrum is linear in $\beta$. In the following sections we study various theoretical and observational implications of the power spectrum (\ref{power-final}). We also comment that our Eq.~(\ref{power-final}) is different from the 
results of Ref. \cite{Cho:2014nka} in which the in-in analysis seems not to be performed rigorously.

Before we end this section, let us discuss one of the most interesting implications of Eq. (\ref{power-final}): the scale-dependence of the domain wall correction to the power spectrum.  Notice that the $\delta$-function in the second term implies that $\bfk_{||}$ = -$\bfq_{||}$. If we now consider the limit $k_z\sim q_z$, Eq. (\ref{power-final}) further implies $k^2\sim q^2$. In this case, the second (asymmetric) term in the equation will behave as $1/k$ compared to the first (isotropic) term. In other words, the asymmetric part will decay on small scales (large $k$). We can go beyond this limit by considering $k_z \gg q_z$. Eq. (\ref{power-final}) is symmetric under $k \leftrightarrow q$ so we do not need to additionally consider the opposite limit. In this case we also find that the domain wall contribution to the power spectrum decays as $1/k$. The only subcase that presents something of an exception is when $k_z \gg k_{||}$ in addition to $k_z \gg q_z$. In that case, we still have $1/k$ decay, but the asymmetric part of the power spectrum has an amplitude which is large (by the same amount that $q_z$ and $k_{||}$ are small). It therefore appears that the $1/k$ decay is general. This has interesting implications in terms of CMB observables and the scale-dependence of the amplitude of the observed CMB power asymmetry: a $1/k$ decay implies a $1/\ell$ decay, where $\ell$ are various multipoles on the CMB sky ($kD \approx \ell$, where $D$ is the distance to the surface of last scattering). The new contributions to the primordial power spectrum from the domain wall therefore imply that the predicted asymmetric features in the CMB must be scale-dependent, which is indeed suggested by observations. Clearly, in order to rigorously study this implication of our model using observational CMB data, we will need to derive the full CMB angular covariance matrix; this is the subject of section \ref{ang-power}.

\section{Two-point correlation and variance in real space}
\label{corr-general}

\subsection{Theoretical predictions}
\label{corr-theory}

In the previous section we derived the power spectrum of primordial curvature perturbations in the presence of a domain wall and found that the new contributions are asymmetric in Fourier space and scale-dependent. In this section, we examine the two-point correlation function in real space, as well as the interesting special case of the variance angular power spectrum. This will enable us to study the angular dependence of the asymmetry, in particular its structure in terms of dipole, quadrupole, octopole, and higher multipole moments.

In order to do this, let us consider only the contribution of the domain wall, and denote its correction to the real space 
two-point correlation by  $ \delta \langle \delta\phi (\tbf{x},\eta_{e})\delta\phi(\tbf{y},\eta_{e})\rangle$. We have
\begin{align}
\delta \big \langle \delta\phi(\tbf{x},\eta_{e})\delta\phi(\tbf{y},\eta_{e}) \big\rangle &= \frac{1}{(2\pi)^6}\int d^3\tbf{k} d^3\tbf{q}e^{i\tbf{k}\cdot\tbf{x}}e^{i\tbf{q}\cdot\tbf{y}}\delta \langle \delta\phi(\tbf{k})\delta\phi(\tbf{q})\rangle  \nonumber \\
&=-\frac{\beta H^2}{4 (2 \pi)^3}\int dk_z dq_z d^2\tbf{q}_{||}\frac{q_{||}^2+k_z q_z}{(q_{||}^2+k_z^2)^{\frac{3}{2}}(q_{||}^2+q_z^2)^{\frac{3}{2}}}e^{i\tbf{q}_{||}\cdot(\tbf{y}_{||}-\tbf{x}_{||}) } \, e^{ik_z  \tbf{x}_3} e^{iq_z  \tbf{y}_3}.
\end{align}
Going to polar coordinates in the $x-y$ plane, $d^2\tbf{q}_{||}=q_{||}  d\theta  dq_{||}$, and using the well-known integral $\int d\theta e^{i  r \cos(\theta)}= 2 \pi J_0(r)$ we get 
\be
\label{power-real}
\delta \big\langle \delta\phi(\tbf{x},\eta_{e})\delta\phi(\tbf{y},\eta_{e}) \big  \rangle = -\frac{\beta H^2}{4(2\pi)^2}\int dk_z dq_zdq_{||} q_{||}\frac{q_{||}^2+k_z q_z}{(q_{||}^2+k_z^2)^{\frac{3}{2}}(q_{||}^2+q_z^2)^{\frac{3}{2}}} J_0(q_{||}|\tbf{y}_{||}-\tbf{x}_{||}|)  \, e^{ik_z  \tbf{x}_3} e^{iq_z  \tbf{y}_3}  .
\ee

It is instructive to look at the \textit{variance} of the curvature fluctuations in real space at any specific point;\footnote{In this subsection we deal with theoretical calculations; the physical significance of the variance is discussed in subsection~\ref{corr-obs}.} this can be obtained by setting $\bfx= \tbf{y}$ in Eq.~(\ref{power-real}):
\ba
\label{var1}
\delta \big \langle \calR^2(\textbf{x})  \big \rangle =
-\frac{\beta }{4}  \calP_0
\int dk_z dq_z dq_{||}q_{||}\frac{q_{||}^2+k_z q_z}{(q_{||}^2+k_z^2)^{\frac{3}{2}}(q_{||}^2+q_z^2)^{\frac{3}{2}}} e^{i(k_z+q_z)z}.
\ea
Note that the ranges of this integral are $0 < q_{||} < \infty$ and $-\infty <  k_z, q_z < \infty$. 
The above expression explicitly shows that the variance is a function only of $z$, the direction perpendicular to the wall. As expected, translational invariance is broken along that direction.  One interesting property of the integral  in Eq.~(\ref{var1}) is that after performing the rescaling 
$q_z \rightarrow z q_z$, $ k_z \rightarrow z k_z$, and $q_{||} \rightarrow z q_{||}$, the integrand becomes independent of $z$. Therefore, any $z$-dependence comes from as how one regularizes the IR divergences for $q_z, k_z, q_{||} \rightarrow 0$.   Performing  the rescaling $q_z \rightarrow z q_z,  k_z \rightarrow z k_z$ and $q_{||} \rightarrow z q_{||}$ the variance in Eq.~(\ref{var1}) becomes
\ba
\delta \big \langle \calR^2(\textbf{x})  \big \rangle = 
-\frac{\beta }{4} \calP_0 
\int_{ 0 }^\infty d q_{||} \int_{-\infty}^\infty d k_z  \int_{-\infty}^\infty d q_z  \frac{
q_{||}^3 \cos(k_z) \cos(q_z) -  q_{||}  k_z q_z \sin(k_z) \sin(q_z) }{ (q_{||}^2 + k_z^2)^{\frac{3}{2}}  (q_{||}^2 + q_z^2)^{ \frac{3}{2}}}.
\ea
One can check that the integral (\ref{var1}) is UV convergent while it is logarithmically divergent in the 
IR region.  To impose the IR cutoff we assume the rescaled momenta satisfy 
$q_{||}, | k_z|, |q_{z}| \ge  |z| /L$, where $L$ is the size of a box encompassing the observable universe \cite{Lyth:2007jh}. Defining 
\ba
F(Q) \equiv   \int_0^\infty \frac{  \cos u  du  }{  (u^2 + Q^2)^{\frac{3}{2}}   }  , \qquad
G(Q) \equiv   \int_0^\infty \frac{  u \sin u  du  }{  (u^2 + Q^2)^{\frac{3}{2}}   }  
=  \int_0^\infty \frac{  \cos u  du  }{  (u^2 + Q^2)^{\frac{1}{2}}   } ,
\ea
and using the symmetry properties of the trigonometric functions, we have
\ba
\label{var-int}
\delta \big \langle \calR^2(\textbf{x})  \big \rangle = - \beta  \calP_0
\int_{|z|/L}^\infty d Q  \left[  Q^3 F(Q)^2 - Q G(Q)^2    \right].  
\ea

As discussed before, the $z$-dependence of the variance comes from the IR cutoff, while it has only a very mild 
dependence on the UV cutoff. One can check numerically that the variance is logarithmically divergent in the IR region. We have checked that, to a very high accuracy, the IR divergence of the variance is given by
\ba
\delta \big \langle \calR^2(\textbf{x})  \big \rangle \simeq 
\beta  \calP_0 \ln \bigg|   \frac{ z}{L}  \bigg| + C,
\ea
where the constant $C$ depends mildly on the UV cutoff.  

\begin{figure}
\centering
\includegraphics[ width=0.4\linewidth]{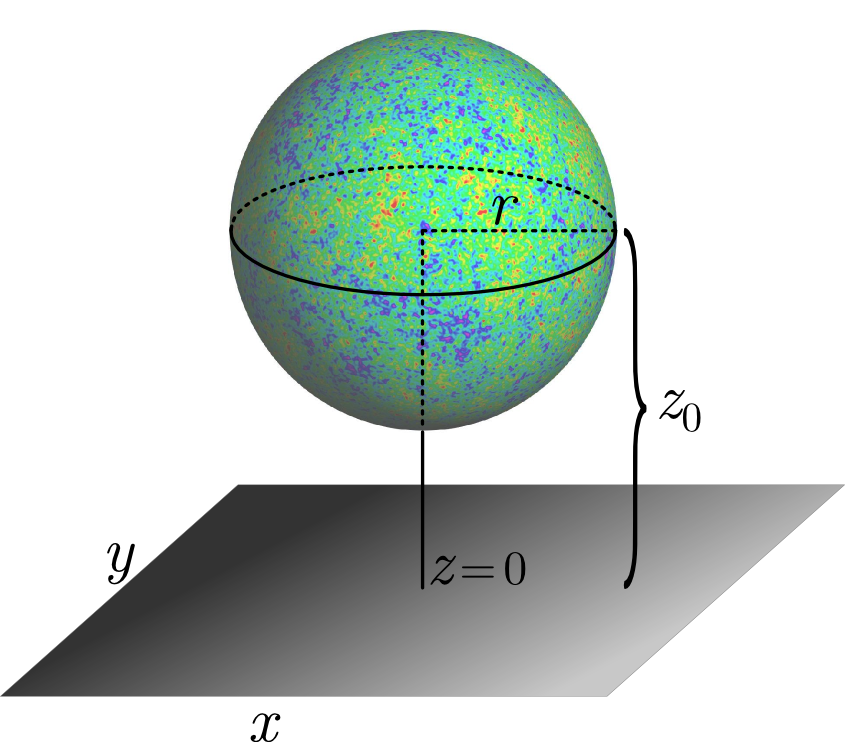}
\includegraphics[ width=0.4\linewidth]{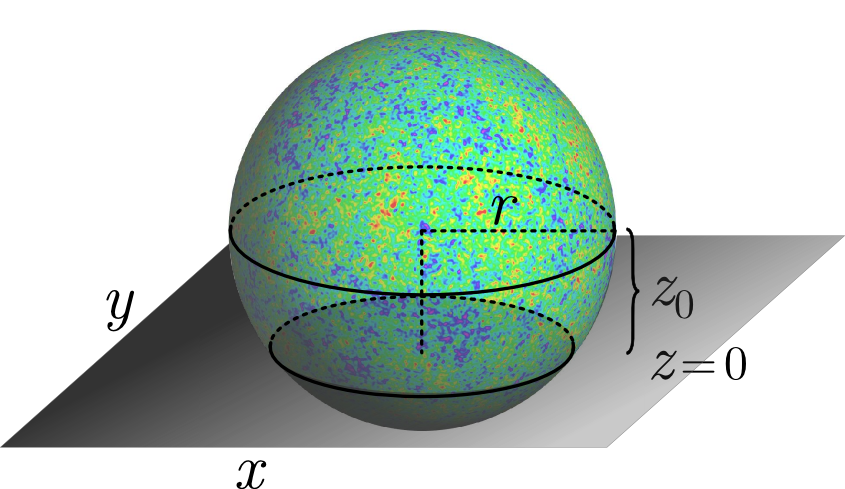}
\caption{The setup for the CMB sphere and the domain wall. {\bf Left:} The case where the CMB sphere does not intersect the domain wall and is located entirely on one side of the wall, corresponding to 
$\kappa \equiv r/z_0 <1$. 
 {\bf Right:} The case where the CMB sphere intersects the domain wall, corresponding to $\kappa>1$.}
\label{fig:cmb-dw}
\end{figure}

To investigate this theory's predictions for the CMB, we consider a two-dimensional sphere 
with a fixed comoving radius $r$ centered  at $z=z_0$. The setup is plotted in Fig.~\ref{fig:cmb-dw}. The center of this CMB sphere is located at
comoving distance $z_0$ from the position of the wall. For any other point on the CMB sphere we have
\ba
z= z_0 + r \cos \theta,
\ea
where $\theta$ is the angle between the point $\bfx$ on the CMB sphere and the 
direction perpendicular to the wall (the $z$ axis). Because of the $\mathbb Z_2$ symmetry of the background geometry, we can consider $z_0  \geq 0$ without loss of generality. This corresponds to the configuration in which the center of the CMB sphere is above the domain wall.  However, we allow for the case where some points on the CMB have $z<0$, i.e., the domain wall intersects the CMB sphere. This occurs when $z_0 <r$, as presented in the right panel of Fig.~\ref{fig:cmb-dw}.  

With this geometrical description in mind, we have
\ba
\label{var-final1}
\delta \big \langle \calR^2(\textbf{x})  \big \rangle \simeq  
\beta  \calP_0  \ln \bigg|   1+ \frac{r}{z_0} \cos \theta  \bigg| + \widehat C,
\ea
where $\widehat C = C - \beta  \calP_0 \ln(z_0/L)$ is a constant which does not affect the physical predictions for the dipole and higher multipoles of the variance on the CMB.

To calculate the dipole ($a_1$), quadrupole ($a_2$), octopole ($a_3$), and higher multipoles for the variance of the curvature perturbations,  
let us decompose the domain wall contribution to the variance in terms of the Legendre polynomials, $P_\ell( \cos \theta )$, as 
\ba
\label{var-expand}
\delta \big \langle \calR^2(\textbf{x})  \big \rangle =    \calP_0 \sum_\ell a_\ell P_\ell(\cos \theta) .
\ea 
Plugging this into Eq.~(\ref{var-final1}) and using the orthogonality condition for the Legendre polynomials, we get
\ba
\label{al_general}
a_\ell = \frac{ ( 2\ell +1 )  \beta}{ 2 } \int_{-1}^{+1} d(\cos \theta) P_\ell(\cos \theta)  \ln \bigg|   1+ \kappa \cos \theta  \bigg|  ,
\ea
where we have defined $\kappa \equiv r/z_0$.  The dipole, quadrupole, and octopole are therefore given by
\ba
a_1 &=& -\frac{3 \beta }{ 4 \kappa^2} \left[  ( \kappa^2 -1 ) \ln\bigg| \frac{1- \kappa}{1+ \kappa} \bigg| -2 \kappa \right],\\
a_2 &=& \frac{ 5 \beta }{ 12 \kappa^3} \left[  3 ( \kappa^2 -1 ) \ln \bigg|  \frac{1- \kappa}{1+ \kappa} \bigg| 
+ 4 \kappa^3 - 6 \kappa  \right],\\
a_3 &=& \frac{7 \beta }{ 48 \kappa^4} \left[  (15 - 18 \kappa^2 + 3 \kappa^4 ) \ln \bigg| \frac{1- \kappa}{1+ \kappa} \bigg|  + 30 \kappa -26 \kappa^3 \right] \, .
\ea

Let us consider the case where $\kappa \ll 1$, i.e., $z_0 \gg r$, so 
the CMB sphere is entirely above the plane of the domain wall and does not intersect it. To lowest order in $\kappa$ we have
\ba
a_1 \simeq \beta \kappa , \qquad
a_2 \simeq -\frac{\beta \kappa^2}{3}  , \qquad
a_3 \simeq \frac{2 \beta \kappa^3}{15} \quad \quad \quad (\kappa \ll 1) .
\label{al_smallkappa}
\ea
We first notice that, because $\beta>0$, our model predicts a negative quadrupole amplitude (note that the sign of the dipole is conventional as $\cos \theta$ changes from the northern hemisphere to the southern one).
Second, the independent parameters of the model are $\beta$ and $\kappa$. 
Eqs. (\ref{al_smallkappa}) then imply that these parameters can be fixed if we measure the dipole and quadrupole amplitudes. This then in turn fixes the
prediction for the octopole amplitude. Indeed, we have 
\ba
| a_3| \simeq \frac{6 a_2^2}{5 | a_1 |}   \quad \quad \quad (\kappa \ll 1) .
\ea
This represents a consistency condition between the dipole, quadrupole, and octopole variance amplitudes.

\begin{figure}[t]
\centering
\includegraphics[ width=0.45\linewidth]{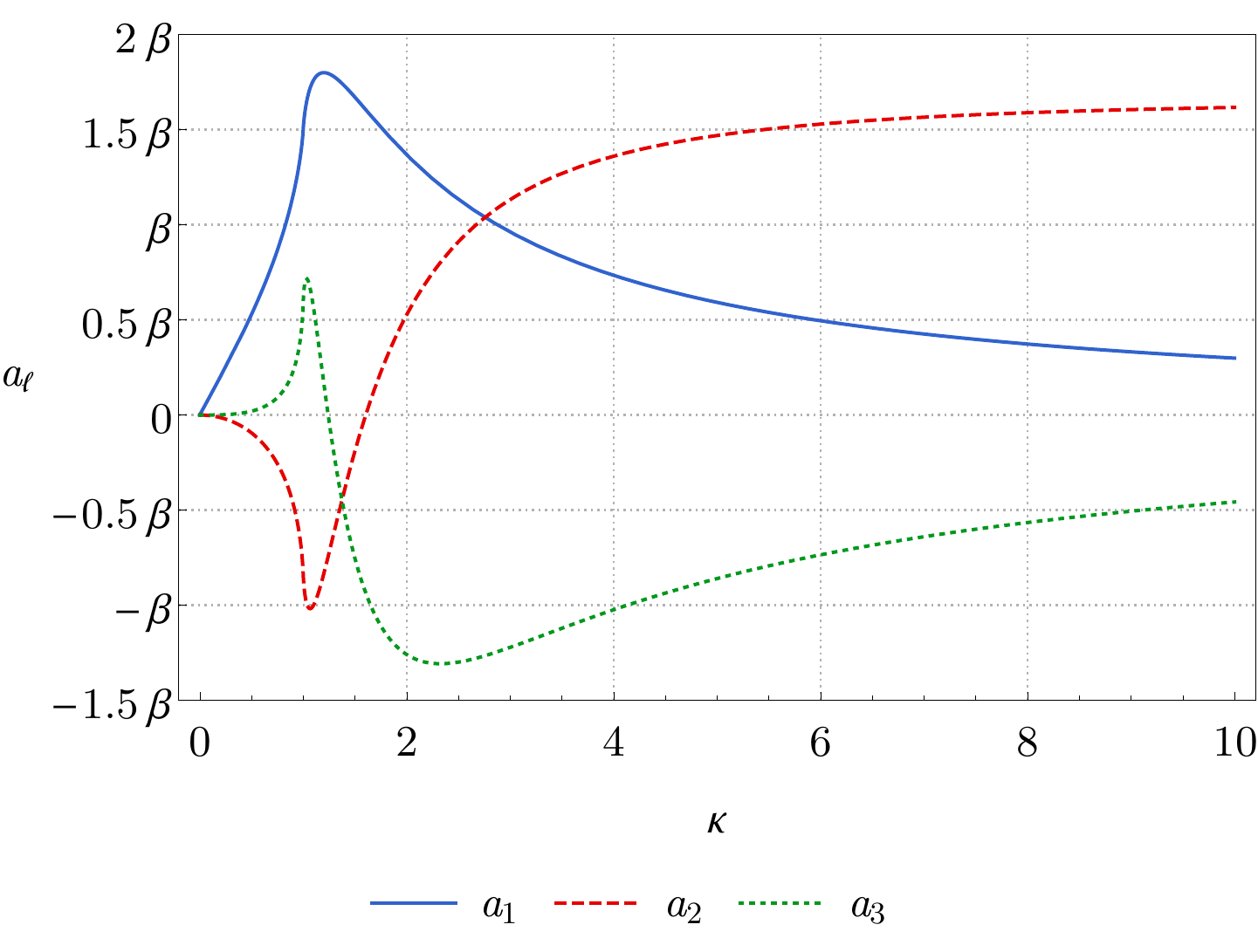}  
\hspace{1cm}
\includegraphics[ width=0.45\linewidth]{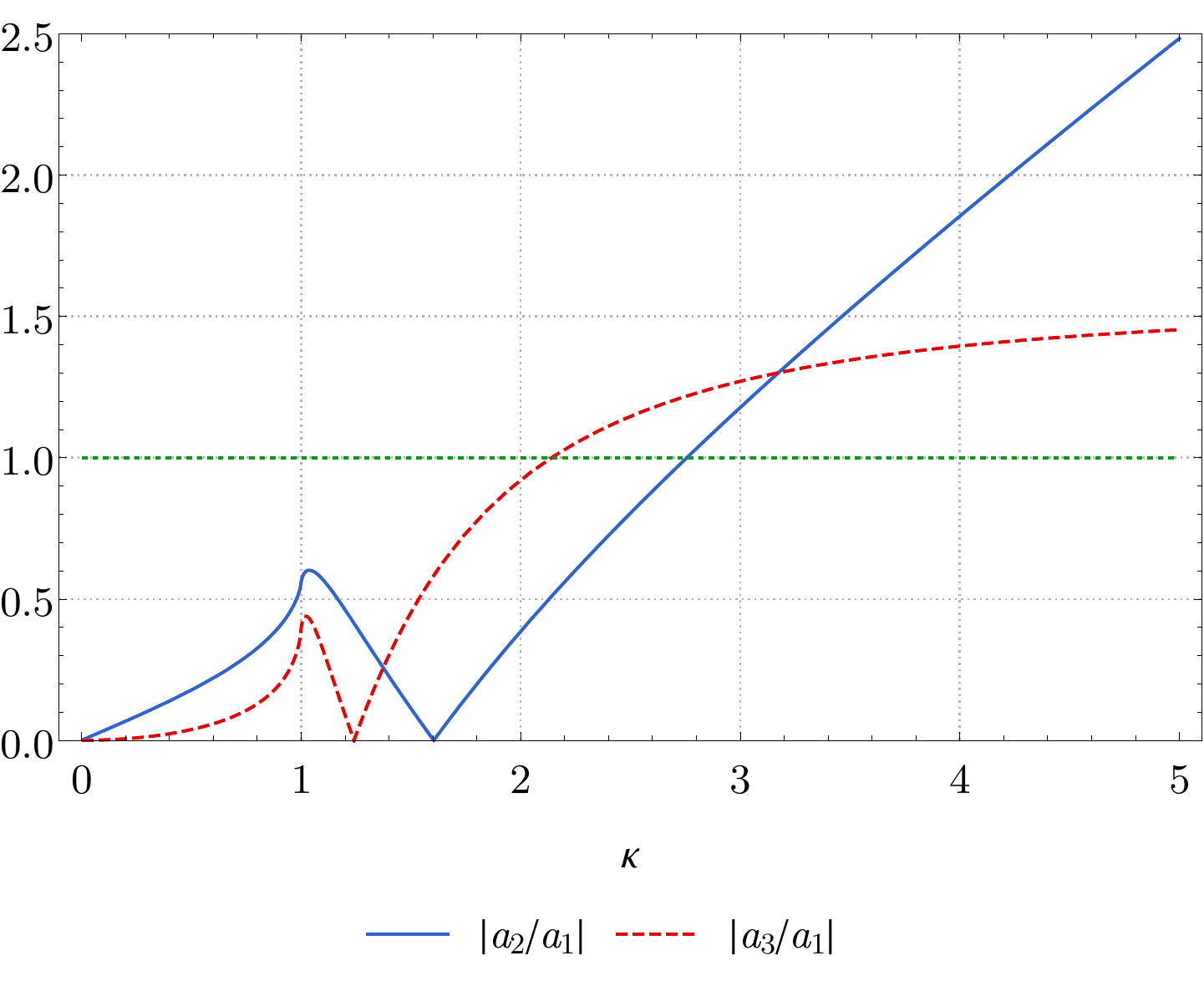}
\caption{{\bf Left:} Values of the variance dipole $a_1$, quadrupole $a_2$, and octopole $a_3$ as functions of $\kappa$, the ratio of the radius of the CMB sphere to our distance to the domain wall. The overall normalization of each multipole is controlled by the parameter $\beta$, which measures the smallness of the domain wall's energy and is taken to be $\ll1$. {\bf Right:} The absolute values of the ratios of the quadrupole $a_2$ and octopole $a_3$ to the dipole $a_1$ as functions of the parameter $\kappa$. Observations suggest $|a_1|\gg |a_2|, |a_3|$ \cite{Akrami:2014eta}; we see this consideration rules out $\kappa\gtrsim2$.}
\label{ai-fig}
\end{figure}

We can further consider the behavior of $a_1$, $a_2$, and $a_3$ for general values of $\kappa$, i.e., when the CMB sphere has an arbitrary position with respect to the plane of domain wall. The behavior of $a_1$, $a_2$, and $a_3$ as functions of $\kappa$ is shown in Fig.~\ref{ai-fig}. Recall that $\kappa>1$ ($\kappa <1$) corresponds to the case in which the domain wall intersects (does not intersect) the CMB sphere. $\beta$ appears as an overall proportionality factor and does not change the behavior of $a_\ell$.

Notice that while $a_1$ is always positive, $a_2$ and $a_3$ change signs depending on the value of $\kappa$. Additionally, in the limit that the domain wall runs through the center of the CMB sphere ($\kappa\to\infty$), the odd multipoles with $\ell =1, 3, \ldots$ fall off like $1/\kappa$, while the even multipoles with $\ell =2, 4,\ldots$ reach constant asymptotic values.
For example, as $\kappa\to\infty$, $a_2\to5 \beta/3$ and $a_4\to-6 \beta/5$. The fact that the odd multipoles vanish in the limit where the center of the CMB sphere is located exactly on the domain wall is expected. This is more easily seen in the case of dipole.  The symmetry considerations imply that there is no dipole when 
$z_0=0$. On the other hand, as $z_0$ deviates from zero and the configuration of the CMB sphere relative to the plane of the domain wall becomes asymmetric, a dipole is expected to develop. If one increases the distance between the center of the CMB sphere and the plane of the domain wall indefinitely, the dipole amplitude is expected to fall off rapidly. Therefore, there exists an intermediate value of $\kappa$ for which the amplitude of dipole reaches a maximum. This conclusion is clearly supported by the left panel of  Fig.~\ref{ai-fig}. Numerically maximizing $a_1$, we find that this maximum occurs at $\kappa \approx 1.2$, corresponding to the domain wall passing near but not through the CMB sphere.

\subsection{Comparison to measured quantities}
\label{corr-obs}

Let us now ask whether the predictions of our model for the multipole amplitudes $a_\ell$ of the variance calculated in the previous section are consistent with measurements of actual CMB sky. In particular, can existing observations place any constraints on the parameters of the model, $\beta$ and $\kappa$?

In order to answer this, let us take a closer look at the definition of the variance, which we calculated in Eq.~(\ref{var1}). Recall that this quantity is just the two-point correlation between different points in space in the limit where the two points are chosen to be identical. Strictly speaking, the correlation function $\big \langle \calR(\textbf{x}) \calR(\textbf{y}) \big \rangle$ and the variance $\big \langle \calR^2(\textbf{x})  \big \rangle$ are ensemble averages of the quantities $\calR(\textbf{x}) \calR(\textbf{y})$ and $\calR^2(\textbf{x})$, respectively, over an infinitely large number of realizations of the Universe. When the two points $\textbf{x}$ and $\textbf{y}$ are located on the CMB sphere (as we always consider), these variances are ensemble averages over different realizations of the CMB.

Since only one realization of the CMB map exists, the variance cannot be measured directly from the data. One can, however, apply some approximations to estimate the value of the variance across the sky. Let us assume, for example, that the statistical properties of the CMB fluctuations (over many realizations of the Universe) do not change significantly in the vicinity of each point on the CMB map. This assumption is justifiable as long as we consider only points that are in close proximity to the selected point. This implies that if we compute the variance of the fluctuations in a very small patch of the sky around a specific point $\textbf{x}$, then that \textit{spatial variance} (or \textit{local variance}) should be a good approximation to the \textit{ensemble variance} $\big \langle \calR^2(\textbf{x})  \big \rangle$. One can compute these local variances across the real CMB sky, at a large number of points, and we may assume that this gives a good approximation to the theoretical ensemble variance that we have calculated for our model.

This procedure has already been performed for both the {\it WMAP} and {\it Planck} CMB maps in Ref.~\cite{Akrami:2014eta}. The authors selected a few thousand points on the CMB, drew a disk of fixed size around each point, and measured the variance within that disk. They then constructed a \textit{local-variance map} which at each point has the variance around that point, and repeated this procedure for a range of disk sizes. They then studied the statistical properties of this variance map, which differ from the standard temperature map.

The authors of Ref.~\cite{Akrami:2014eta} showed that the local-variance map has a statistically significant dipole. More specifically, they computed the local-variance maps for 1000 simulations of isotropic universes, and found that none of these has a variance dipole as large as that in our Universe. A particularly interesting --- and relevant, for our purposes --- result of the analysis in Ref.~\cite{Akrami:2014eta} is the angular power spectrum of the local-variance map presented in their Fig. 2(d). This quantity, called $C_\ell$ in that reference,\footnote{This should not be conflated with the more familiar $C_\ell$ of the temperature angular power spectrum. We are referring to the power spectrum of the \textit{local-variance map}, which at each point contains the temperature variance around that point, rather than the temperature itself.} corresponds directly to $a_\ell^2$, where the $a_\ell$ are defined in Eq.~(\ref{var-expand}). Note that the quantity in Eq.~(\ref{var-expand}) contains only the domain wall's contribution to the variance; however, for a purely isotropic setup (i.e., in the absence of the domain wall) the variance is expected to be independent of the direction on the sky and therefore to comprise only a monopole $a_0$.

The pressing question therefore is: can our anisotropic domain wall model provide a set of $a_\ell$ ($\ell=1,2,3,...$) which are consistent with the results of Ref.~\cite{Akrami:2014eta}? Remarkably, we will find that the answer is \textit{yes}.

\begin{figure}[t]
\centering
\includegraphics[ width=0.5\linewidth]{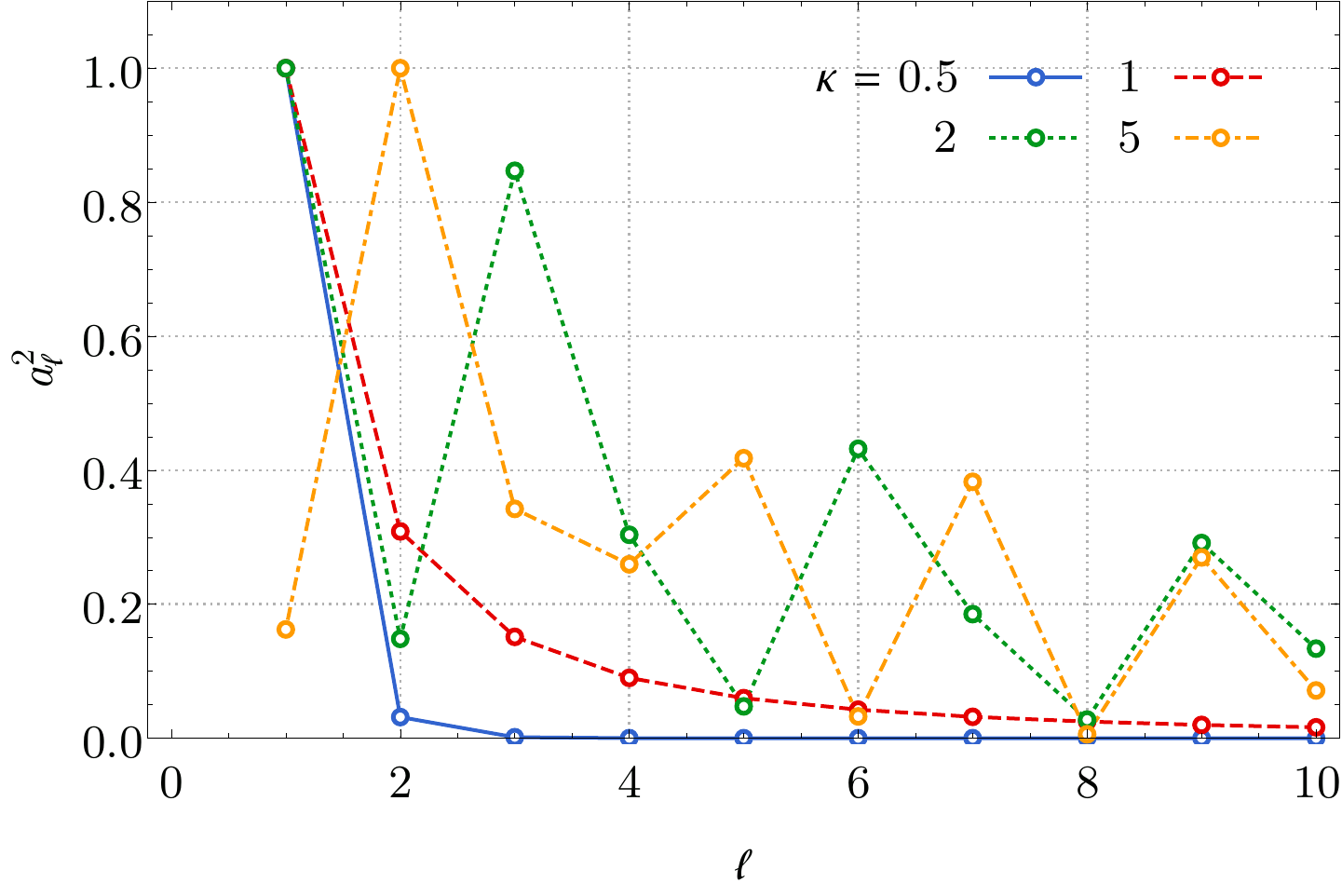}
\caption{The variance angular power spectra $a_\ell^2$ for a range of $\kappa$, where $a_\ell$ are the amplitudes of different variance multipoles ($a_1$, $a_2$, and $a_3$ correspond to the dipole, quadrupole and octopole, respectively). The magnitude of $a_\ell^2$ changes significantly with $\kappa$, so for comparison each line has been normalized so that the largest multipole is equal to unity. We do not plot $\kappa<0.5$ because, while the amplitude decreases, the shape effectively stays the same. It is interesting to note that, as soon as $\kappa$ climbs above unity, the $a_\ell^2$ begin to oscillate (we have checked that this behavior emerges even for $\kappa$ as low as 1.1). These plots can be directly compared to measured data in Fig. 2(d) of Ref.~\cite{Akrami:2014eta}; the cases with $\kappa\lesssim1$ agree with the shape of the observed spectrum quite well (see Fig.~\ref{al-chi2-fig}).}
\label{al2-fig}
\end{figure}

We begin by plotting, in Fig.~\ref{al2-fig}, our predictions for the variance angular power spectrum $a_\ell^2$ for four different values of $\kappa$, calculated using Eq.~(\ref{al_general}). This corresponds directly to the observed power spectrum in Fig. 2(d) of Ref.~\cite{Akrami:2014eta}. Note that the values are normalized to the largest $a_\ell^2$ for each case, in order to allow us to directly compare the curves' shapes, rather than their amplitudes. This figure shows that for small values of $\kappa$, i.e., for cases where the domain wall does not intersect the CMB sphere, the dipole contribution $a_1$ is dominant over all higher multipoles ($\ell=2,3,...$), and $a_\ell^2$ decreases monotonically with increasing $\ell$. This is consistent with what has been observed in the real data \cite{Akrami:2014eta}.

The CMB variance map in Ref.~\cite{Akrami:2014eta} displays a large dipole, while all the other multipoles are small and consistent with zero. As a result,  large values of $\kappa$ --- $\kappa\gtrsim2$ --- 
are ruled out, as can be seen from both Figs.~\ref{ai-fig} (right panel) and \ref{al2-fig}. Above this value, the quadrupole and octopole grow large compared to the dipole, which is in clear disagreement with Fig.~2(d) of Ref.~\cite{Akrami:2014eta}.\footnote{The dipole is the dominant contribution to the variance power spectrum for $0\leq\kappa<2.14$. The octopole then dominates for $2.14<\kappa<3.17$. Thereafter the quadrupole is always the largest multipole.} We can therefore rule out the model for $\kappa\gtrsim2$. This means that in order for our model to produce the observed variance asymmetry, the center of the CMB sphere cannot be too close to the domain wall: the wall must either lie entirely outside the CMB sphere, or should pass no closer to its center than about half the sphere's radius. This is a conservative bound; even at $\kappa=2$, the quadrupole amplitude in Fig.~\ref{al2-fig} is likely too large to be in good agreement with the data. We will now discuss further constraints.

There are two competing effects which could, in principle, spell trouble for our model. We need $\kappa$ to be small in order to produce the mostly-dipole variance that is suggested by observations. However, the lower $\kappa$ is, the smaller the amplitude of the dipole. The amplitude of $a_\ell$ is proportional to $\beta$, but $\beta$ (which measures the energy of the domain wall relative to the dominant inflaton potential) is assumed to be a small parameter. Therefore we could potentially run into issues in achieving both a large enough dipole and small enough higher multipoles.

There is another potential problem in that we have measured several multipole moments in the CMB variance map, and only have two free parameters ($\beta$ and $\kappa$) with which to fit them. To see whether there are regions of parameter space which agree with observations, and further elucidate the constraints (beyond the heuristic $\kappa\lesssim2$ which we have discussed above), we now compare the predictions directly to the observed multipole amplitudes. This can be done by constructing a likelihood for our model based on the measured multipole amplitudes given in Fig.~2(d) of Ref.~\cite{Akrami:2014eta}. In order to do this, we first need the measured values and their associated uncertainties.

The results of Ref.~\cite{Akrami:2014eta} show that for $6^{\circ}$ disks, which seem to give optimal measurements of the multipole amplitudes, the measured value for $|a_1|$ (the absolute value of the local-variance dipole amplitude) is $\sim 0.03$ with a $1\sigma$ error of $\sim 0.007$ (for the \textit{Planck} temperature data).\footnote{The measured values for $a_\ell$ given in Fig.~2(d) of Ref.~\cite{Akrami:2014eta} are somewhat different from the ones we use in this paper. We have corrected those values for the deviations from zero in the distributions of the amplitudes for isotropic simulations. These deviations are believed to be artifacts of the masked sky maps used in the analysis of Ref.~\cite{Akrami:2014eta}.} For higher multipoles, the amplitudes are consistent with zero and can have both positive and negative values. We use the amplitudes for multipoles up to $\ell=5$ in our analysis; we have checked that including higher multipole moments does not change our results significantly. In summary, we have the measured values
\ba
|a_{1,\mathrm{obs}}|=0.03,\qquad |a_{2,\mathrm{obs}}|=0,\qquad |a_{3,\mathrm{obs}}|=0,\qquad |a_{4,\mathrm{obs}}|=0,\qquad |a_{5,\mathrm{obs}}|=0,
\ea
and the associated (conservative) uncertainties\footnote{The grey points in Fig. 2(d) of Ref.~\cite{Akrami:2014eta} are the amplitudes obtained from isotropic simulations of the CMB map, and their distribution at each multipole provides the probability that the null hypothesis is true, i.e., that the amplitude of the variance at that multipole is zero (or consistent with isotropy). These distributions therefore cannot be used to infer the exact uncertainties around the measured values of $a_\ell$ for our analysis and provide only rough approximations to the values. We leave a rigorous statistical analysis of the model for future.}
\ba
\sigma_1=0.007,\qquad \sigma_2=0.01,\qquad \sigma_3=0.01,\qquad \sigma_4=0.008,\qquad \sigma_5=0.008.
\ea

Using these estimated measurements and errors, we construct
\begin{equation}
\chi^2 = \displaystyle\sum_{\ell=1}^5 \left(\frac{|a_{\ell,\mathrm{pred}}| - |a_{\ell,\mathrm{obs}}|}{\sigma_\ell}\right)^2
 \label{eq:chi2}
\end{equation}
over the parameter space with $\beta<0.1$ (in order for our perturbative approach to be valid) and $\kappa<2$ (so that the dipole is dominant). Note that we take the absolute value of the predicted $a_\ell$ as we only measure $a_\ell^2$, even though $a_\ell$ is sometimes negative (see Fig.~\ref{ai-fig}).

\begin{figure}[t]
\centering
\includegraphics[ width=0.6\linewidth]{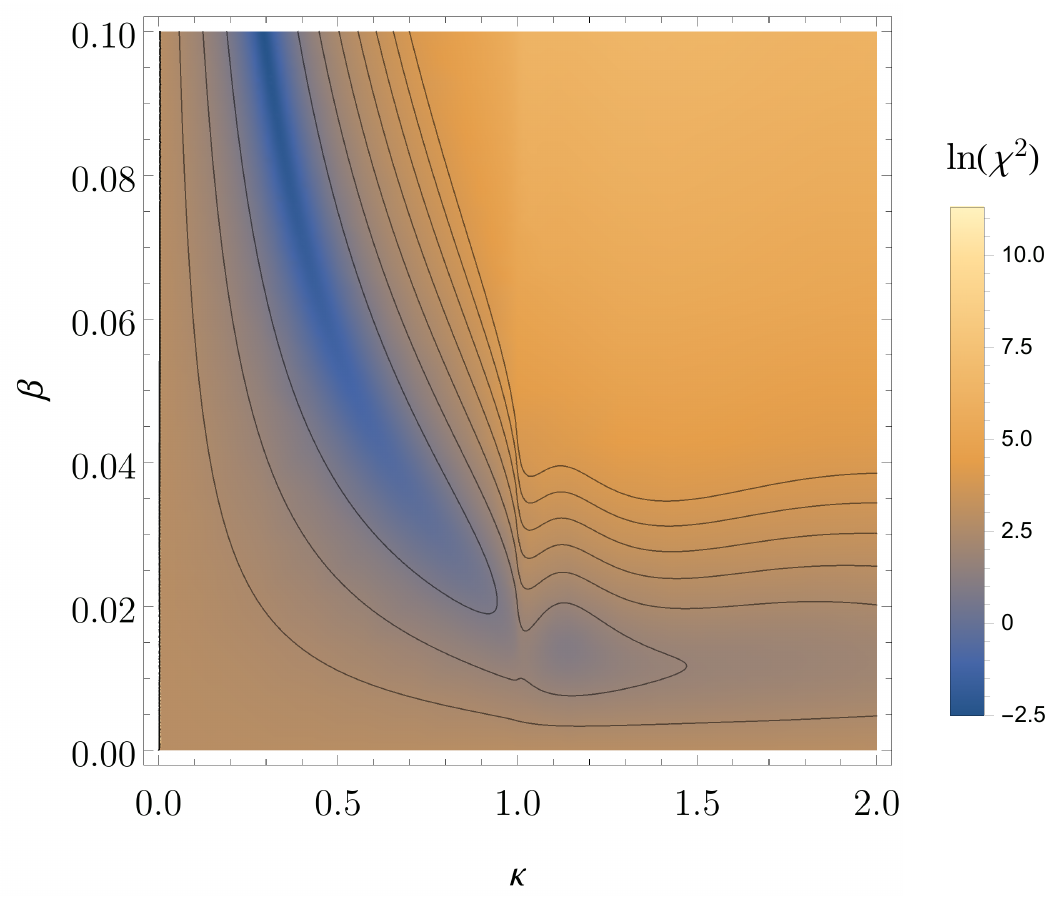}
\caption{Here we plot $\chi^2$, defined in Eq.~(\ref{eq:chi2}), for $\beta<0.1$ and $\kappa<2$. This measures how much the predicted dipole and higher multipoles deviate from the values measured in Ref.~\cite{Akrami:2014eta}. The contour lines correspond to $1\sigma$, $2\sigma$, etc. deviations from the best-fit point. More precise constraints on $\beta$ and $\kappa$ will be possible when the errors for $a_\ell$ are measured more accurately.}
\label{al-chi2-fig}
\end{figure}

A density-contour plot of $\chi^2$ is shown in Fig.~\ref{al-chi2-fig}. The first question is whether any points in this space fit the data, and we see that they do. The color of a point corresponds to its $\chi^2$, and the darker blue regions have small $\chi^2$, which suggests a good fit to observations. The second question is, given that some points fit well, what constraints we can put on the parameter space. This is measured out by the contours, which are the sets of points $1\sigma$, $2\sigma$, etc. away from the best-fit point (i.e., the point with the lowest $\chi^2$). The best-fit point is $(\beta,\kappa)=(0.1,0.29)$, which has $\chi^2_\mathrm{min}=0.09$.\footnote{The $1\sigma$, $2\sigma$, etc. contours correspond to $\chi^2-\chi_\mathrm{min}^2=(2.3, 6.18, 11.83, 19.33, 28.74, 40.09, 53.38)$, up to $7\sigma$.}

The contours indicate that small values of $\kappa$ and large values of $\beta$ are favored; this can be seen from the banana-shaped $1\sigma$ region of Fig.~\ref{al-chi2-fig}. This result is not surprising; our analytical discussions have already told us (see Fig.~\ref{al2-fig}) that for values of $\kappa$ smaller than unity (i.e., when the CMB sphere does not intersect the domain wall) the dipole amplitude is dominant over other multipoles and $|a_\ell|$ monotonically decreases with increasing $\ell$ (multipole amplitudes for $\ell>1$ are significantly smaller than the dipole amplitude). This agrees with the results of Ref.~\cite{Akrami:2014eta}. However, making $\kappa$ small reduces the magnitude of the dipole $a_1$. The magnitude also depends (linearly) on $\beta$, so the fairly large observed value of $a_1$ prefers higher values of $\beta$. It is non-trivial, and a good sign for this model, that we can get a large enough amplitude while keeping $\beta$ a perturbative parameter. So in this region, the predicted value of $a_1$ will be large and close to the measured one while the model will give arbitrarily small values for $|a_\ell|$ ($\ell>1$), i.e., values that are consistent with zero.

There is also a $2\sigma$ region with $1\lesssim\kappa\lesssim1.5$ in Fig.~\ref{al-chi2-fig}. This is a reflection of the fact that $a_1$ is very large at those values of $\kappa$ even when $\beta$ is quite small. In this region the dipole is still dominant over the other multipoles, as can be seen from both panels of Fig.~\ref{ai-fig} for the quadrupole and octopole. The reason why the points in that region do not fit the data as well as the small-$\kappa$ and large-$\beta$ region is that $|a_\ell|$ oscillates with $\ell$ when $\kappa>1$. Consequently, $|a_4|$ and $|a_5|$ become unacceptably large, even though $|a_2|$ and $|a_3|$ remain very small (consistent with zero). We performed a similar $\chi^2$ analysis leaving out $a_4$ and $a_5$, and many points within the large-$\kappa$ $2\sigma$ region of Fig.~\ref{al-chi2-fig} formed a $1\sigma$ island, not connected to the ``main'' $1\sigma$ region. This vanishes when $a_4$ and $a_5$ are included, demonstrating the importance of including higher multipoles in the fit to data.

In summary, based on a rough statistical analysis, in which we compared our predictions for the structure of the local variance (as illustrated in Fig.~\ref{al2-fig}) to its analogue from the real data (as presented in Fig. 2(d) of Ref.~\cite{Akrami:2014eta}), we conclude that our model is able to produce the structure of the anomalous asymmetry observed in the actual CMB sky, where the dipole contribution is large and dominant over higher multipoles. This is consistent with previous reports on the success of fitting a simple, phenomenological dipole modulation model to the large-scale (low-$\ell$) CMB fluctuations. In addition, this seems to favor small values of $\kappa$, because for large values the dipole is not dominant, contrary to observations. We conclude that a domain wall lying entirely outside the CMB sphere and with an energy density (over a Hubble volume) subdominant to the inflaton potential by about an order of magnitude is a consistent and simple explanation for the observed power asymmetry in the CMB. The structure of the asymmetry in our model is more sophisticated than a dipole modulation and an extensive statistical analysis, both in real space and harmonic space, is required to test the model more appropriately and further constrain its parameters; we leave this for future work.

\section{CMB angular power spectra}
\label{ang-power}

In this section, we calculate the corrections to the angular power spectra from the domain wall anisotropies. This is a natural next step since those are the quantities that are used in standard comparisons of the predictions of a model to the CMB data \citep[see, e.g.,][]{Ade:2013kta}. In the standard isotropic cosmological model, the covariance matrix of the primordial CMB fluctuations in spherical harmonic space is diagonal and given solely by the CMB power spectrum $C_\ell$. Any statistically anisotropic feature produces non-diagonal elements and one therefore needs to know the full covariance matrix. Our anisotropic model is no exception and we expect it to also predict a non-trivial covariance matrix; the aim of this section is to derive explicit expressions for different elements of that matrix. In the following, we may refer to the elements of the covariance matrix simply as angular power spectra.

Let us start by expanding the comoving curvature perturbations in terms of spherical harmonics. We have
\begin{align}
\label{alm}
a_{\ell m} = 4\pi  \int \frac{d^3k}{(2\pi)^3} \Delta_\ell(k) \mathcal{R}_\mathbf{k} Y_{\ell m}^*(\hat k), 
\end{align} 
where $\Delta_\ell(k)$ is the radiation transfer function, which can be calculated numerically with a Boltzmann code; we use CLASS \cite{Blas:2011rf} to produce our numerical results in this paper. 

To calculate the covariance matrix $\langle a_{\ell_1, m_1} a_{\ell_2, m_2}\rangle$, we make use of the primordial correlation function, Eq.~(\ref{power-final}). Note that Eq.~(\ref{power-final}) implicitly chooses the location of the domain wall to be $z=0$. However, for the purposes of this calculation, it is convenient to choose the center of the CMB sphere to be located at the origin. The shift of spatial coordinates corresponds to multiplying the RHS of Eq.~(\ref{power-final}) by $e^{i(k_z+q_z)z_0}$, where $z_0$ is the distance between the domain wall and the center of the CMB sphere. As a result, the two-point correlation function of $a_{\ell m}$ can be written as
\begin{align}
\label{covmat}
\langle a_{\ell_1, m_1} a_{\ell_2, m_2}\rangle = 
4\pi 
\mathcal{P}_0 \delta_{\ell_1, \ell_2} \int \frac{dk}{k}~\Delta_{\ell_1}^2(k) 
-
(4\pi)^2 \mathcal{P}_0
2\pi^2  \beta ~ \mathcal{T}(\ell_1, \ell_2, m_1, m_2),
\end{align}
where the first term is the standard isotropic contribution, and $ \mathcal{P}_0 = \left(\frac{H^2}{2\pi \dot\phi}\right)^2$ is the scale-invariant primordial power spectrum. The contribution from the domain wall takes the form
\begin{align}
  \mathcal{T}(\ell_1, \ell_2, m_1, m_2) 
  \equiv  \pi \int_{-\infty}^\infty \frac{dq_1}{2\pi}~ & \int_{-\infty}^\infty \frac{dq_2}{2\pi}~ 
  \int_0^\infty \frac{d^2p}{(2\pi)^2}~ 
  (p^2 + q_1 q_2) e^{i(q_1-q_2)z_0} \times
\nonumber\\ &
  \frac{\Delta_{\ell_1}(\sqrt{p^2+q_1^2})Y_{\ell_1, m_1}^*(\cos\theta_1)}{(p^2+q_1^2)^{3/2}}~
  \frac{\Delta_{\ell_2}(\sqrt{p^2+q_2^2})Y_{\ell_2, m_1}^*(\cos\theta_2)}{(p^2+q_2^2)^{3/2}},
\end{align}
where $\theta_i\equiv \arctan (p/q_i)$.

Note that the domain wall does not break the rotational symmetry along the $z$-axis. Thus we should be able to integrate out the angular direction along the domain wall. This can be done by reducing the spherical harmonics into the associated Legendre polynomials $P_{\ell}^{m}$ and integrating out the angle $\phi$:
\begin{align}
  \int d\phi  Y_{\ell_1, m_1}^*(\theta_1, \phi) Y_{\ell_2, m_2}^*(\theta_2, \phi)
  = 
  2\pi \delta_{m_1, m_2}
  \sqrt{\frac{(2 \ell_1 + 1)}{4\pi} \frac{(\ell_1-m_1)!}{(\ell_1+m_1)!}}    
  P_{\ell_1}^{m_1}(\cos\theta_1)
  \sqrt{\frac{(2 \ell_2 + 1)}{4\pi} \frac{(\ell_2-m_2)!}{(\ell_2+m_2)!}}    
  P_{\ell_2}^{m_2}(\cos\theta_2).
\end{align}
This immediately tells us that $\mathcal{T}=0$ when $m_1\neq m_2$. For $m_1=m_2=m$, we find
\begin{align}\label{eq:calc}
  \mathcal{T} & (\ell_1, \ell_2, m_1, m_2) = \mathcal{T}(\ell_1, \ell_2, m, m) =
 \frac{\sqrt{(2 \ell_1+1)(2 \ell_2+1)}}{8\pi^3}  
  \sqrt{\frac{(\ell_1-m)!}{(\ell_1+m)!}}
  \sqrt{\frac{(\ell_2-m)!}{(\ell_2+m)!}} ~\times
  \nonumber\\&
  \left[
  \int_0^\infty  p^3 dp ~
  \int_0^\infty  dq_1~ \cos(q_1z_0)
  P_{\ell_1}^{m}(\cos\theta_1)
  \frac{\Delta_{\ell_1}(\sqrt{p^2+q_1^2})}{(p^2+q_1^2)^{3/2}} ~
  \int_0^\infty  dq_2~ \cos(q_2z_0)
  P_{\ell_2}^{m}(\cos\theta_2)
  \frac{\Delta_{\ell_2}(\sqrt{p^2+q_2^2})}{(p^2+q_2^2)^{3/2}}
  \right.
  \nonumber\\ +&
  \left.
  \int_0^\infty  pdp ~
  \int_0^\infty  q_1 dq_1~ \sin(q_1z_0)
  P_{\ell_1}^{m}(\cos\theta_1)
  \frac{\Delta_{\ell_1}(\sqrt{p^2+q_1^2})}{(p^2+q_1^2)^{3/2}} ~
  \int_0^\infty  q_2 dq_2~ \sin(q_2z_0)
  P_{\ell_2}^{m}(\cos\theta_2)
  \frac{\Delta_{\ell_2}(\sqrt{p^2+q_2^2})}{(p^2+q_2^2)^{3/2}}
  \right].
\end{align} 
By numerically computing $\mathcal{T}$ for any choices of $\ell_1$, $\ell_2$, and $m$ we can compute the entire covariance matrix for any chosen values of $z_0$ and $\beta$. Note that the dependence on the comoving radius of the CMB sphere (or our distance to the last scattering surface) is implicit in these expressions by how the momenta $\mathbf{p}$ and $q_i$ correspond to the multipoles $\ell_i$. This means that the parameter $z_0$ here plays the same role as the parameter $\kappa$ used in previous sections.

Let us now compute some elements of the covariance matrix (\ref{covmat}) for specific values of $\ell_1$, $\ell_2$, $m$, and $z_0$ to examine their dependence on these quantities. We carry out all the integration in Eq.~(\ref{eq:calc}) numerically.

\begin{figure}[t]
\centering
\includegraphics[ width=0.45\linewidth]{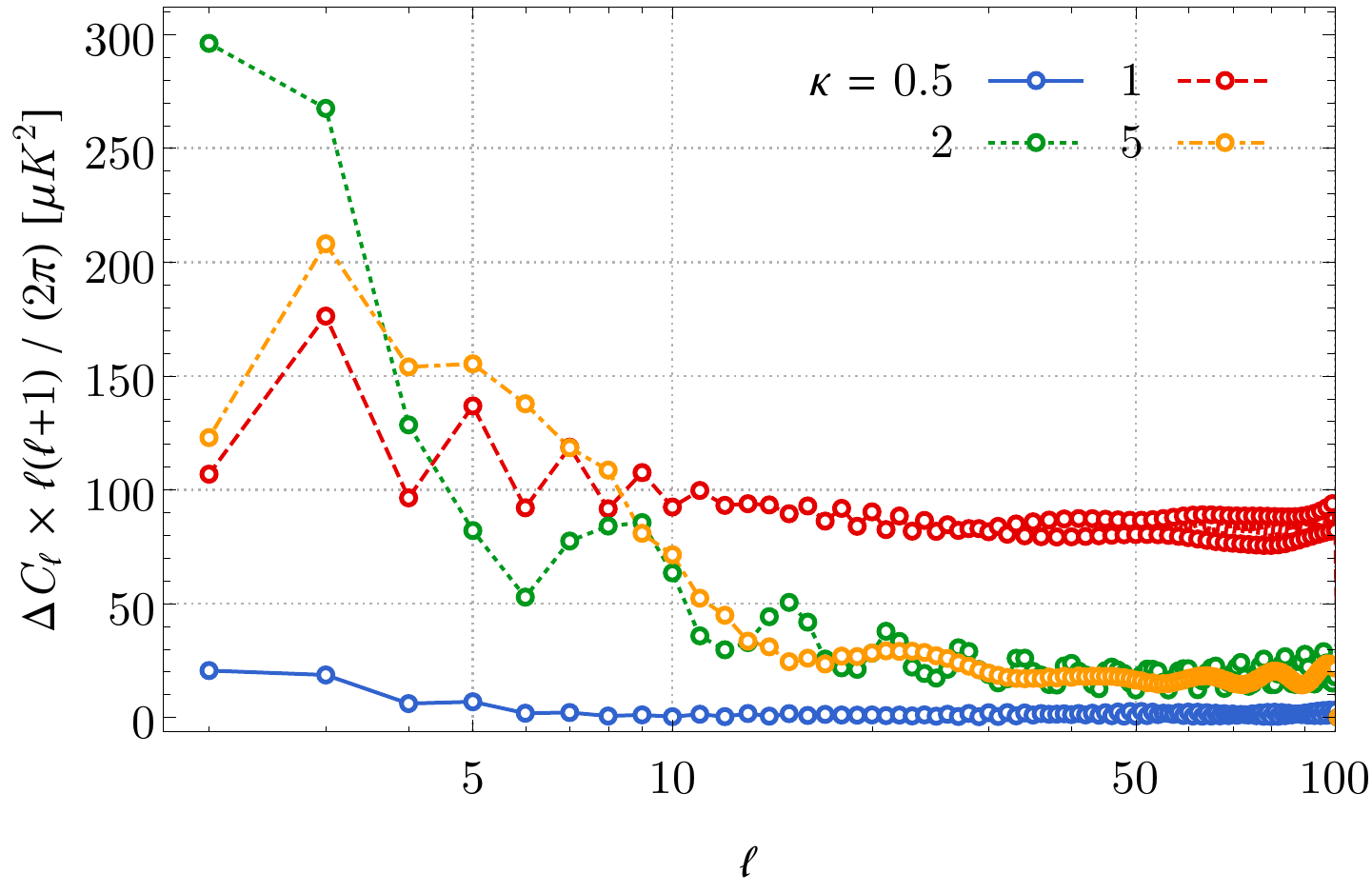}  
\hspace{1cm}
\includegraphics[ width=0.45\linewidth]{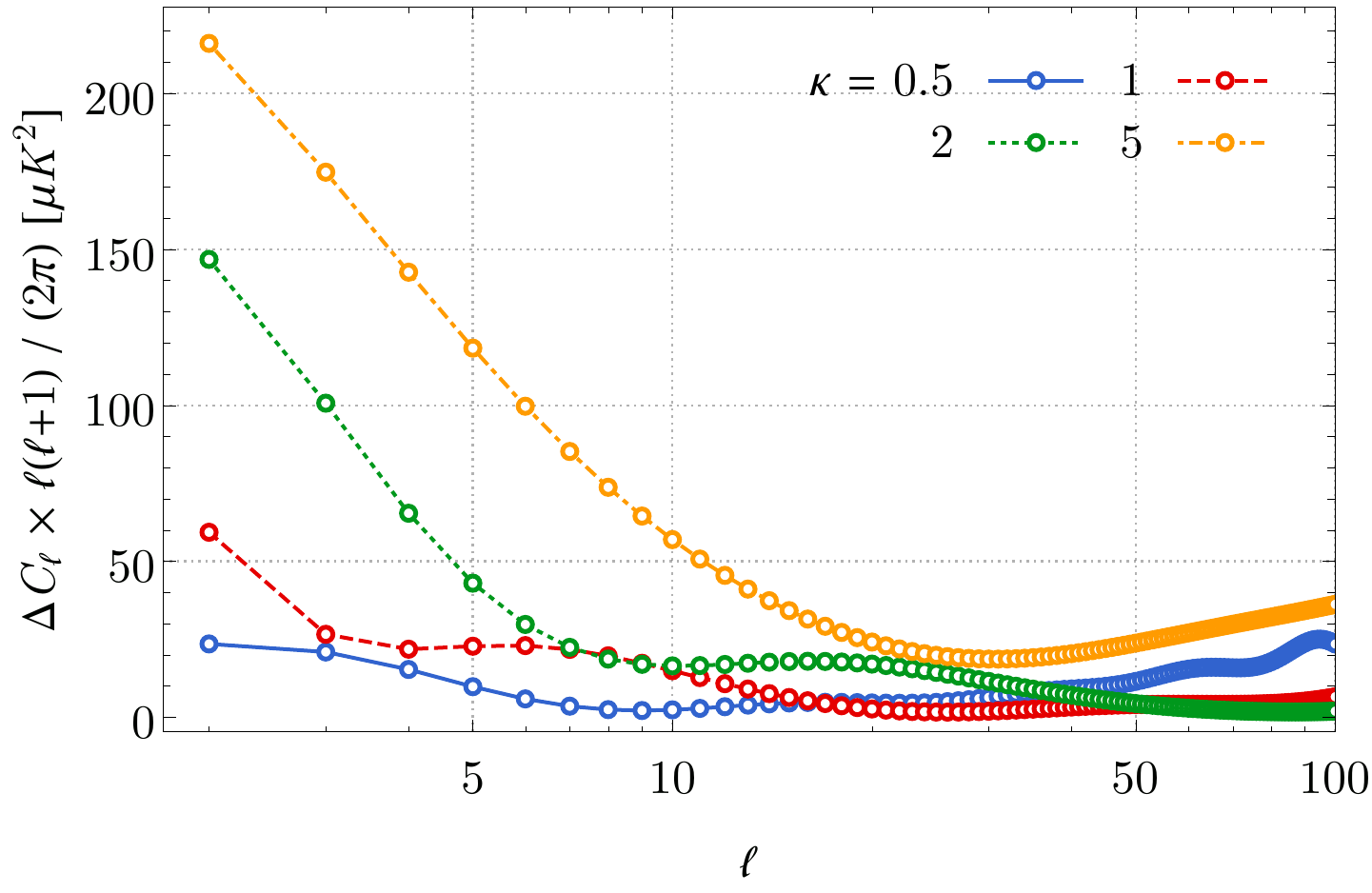}
\caption{Contributions of the domain wall to the CMB power spectrum $C_{\ell}$ for different multipoles and for four values of the parameter $\kappa$. {\bf Left:} We set $m=0$. {\bf Right:} We set $m=\ell$ at each multipole $\ell$. In both panels the plotted quantities correspond to $\langle a_{\ell_1, m} a_{\ell_2, m}\rangle$ in Eq.~(\ref{covmat}), where we set $\ell_1=\ell_2=\ell$.}
\label{dell0m0mell-fig}
\end{figure}

In Fig.~\ref{dell0m0mell-fig}, we show the contribution of the domain wall to the CMB power spectrum $C_\ell$ for the cases where $\ell_1$ and $\ell_2$ are identical, i.e., the correction to the diagonal elements of the covariance matrix
\begin{align}
  \Delta C_{\ell} \equiv 
-(4\pi)^2 \mathcal{P}_0  2\pi^2  \beta ~ \mathcal{T}(\ell, \ell, m, m).
\end{align}
The left panel presents the quantities for $m=0$ while the right panel corresponds to cases where $m$ has been set to $\ell$ at each $\ell_1=\ell_2=\ell$. In each panel, we present the results for four different configurations of the CMB sphere and the domain wall, parameterized by $\kappa=0.5$, 1, 2, and 5. The value of $\beta$ is set to unity in both panels, as $\beta$ only provides an overall scaling.

Fig.~\ref{dell0m0mell-fig} implies that:
\begin{enumerate}
\item In each case, the domain wall makes larger contributions to the power spectra on large scales, and the contributions decay with increasing $\ell$, although there are oscillations and slight enhancements in certain cases.

\item Increasing our distance to the domain wall decreases the extra power added by the domain wall, as expected: the closer the CMB sphere is to the domain wall, the stronger the effects of the wall are on the power. Based on our discussions on the variance in the previous section, only the cases with $\kappa=0.5$ and $1$ (amongst the four cases shown in the plots) are in good agreement with observations: they produce strong dipole and weak higher multipole asymmetries. For these cases the contributions to the power spectrum are relatively small.
\end{enumerate}
We also checked a few cases where the power spectra were computed with $\ell_1\neq\ell_2$ and the results showed similar behavior; we do not present them here for brevity.

\begin{figure}[t]
\centering
\includegraphics[ width=0.46\linewidth]{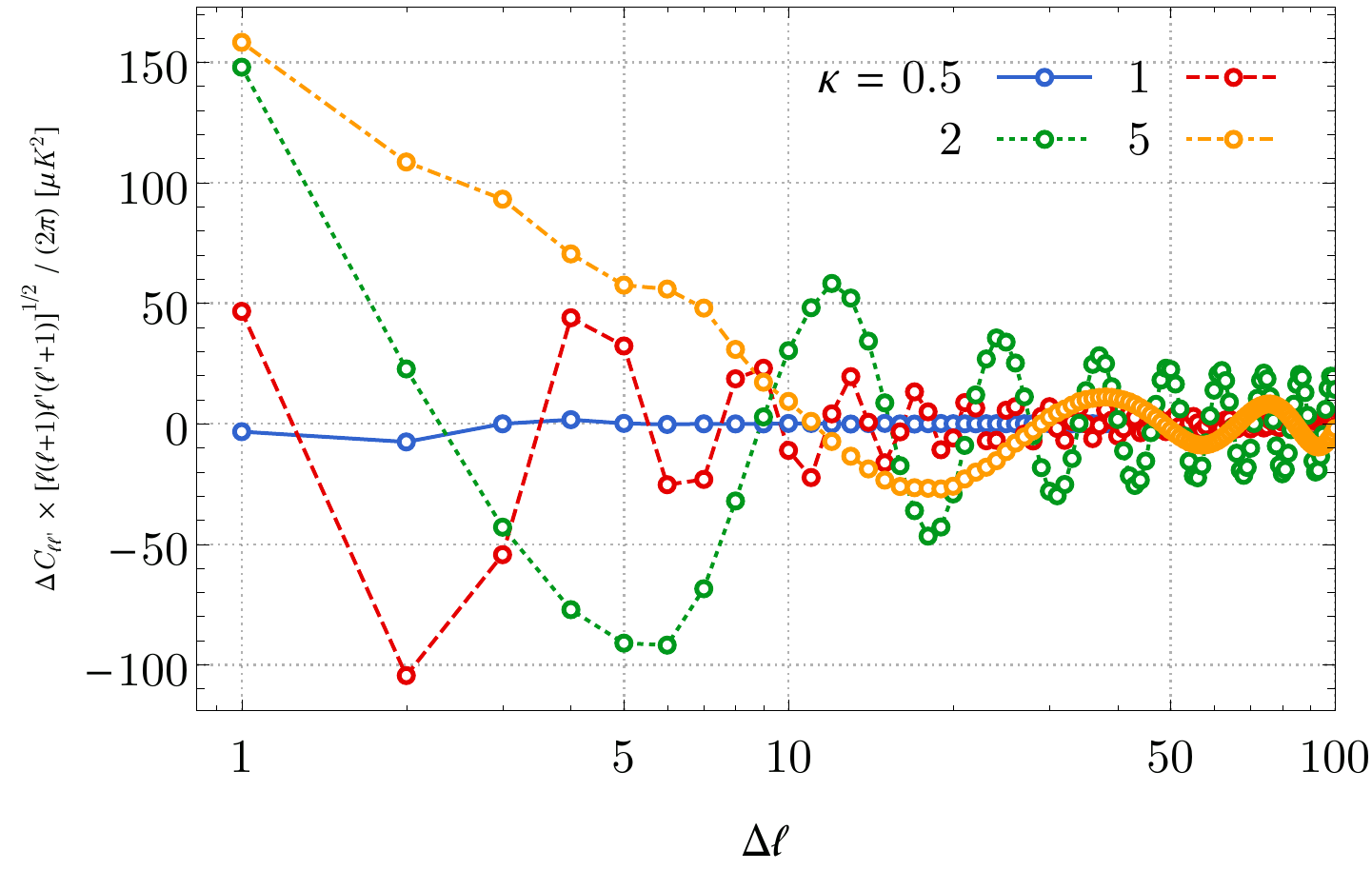}  
\hspace{1cm}
\includegraphics[ width=0.46\linewidth]{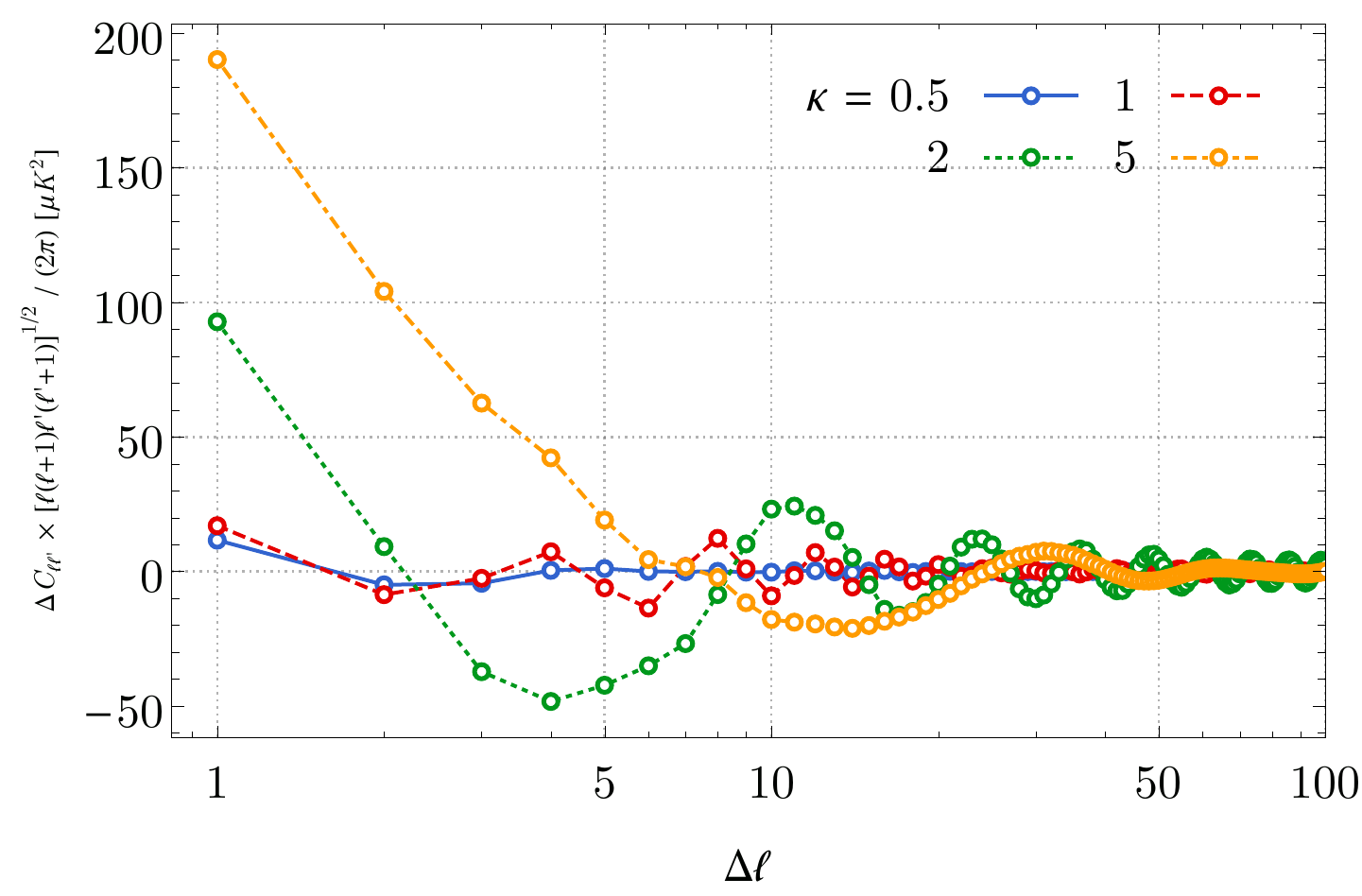}
\hspace{1cm}
\includegraphics[ width=0.46\linewidth]{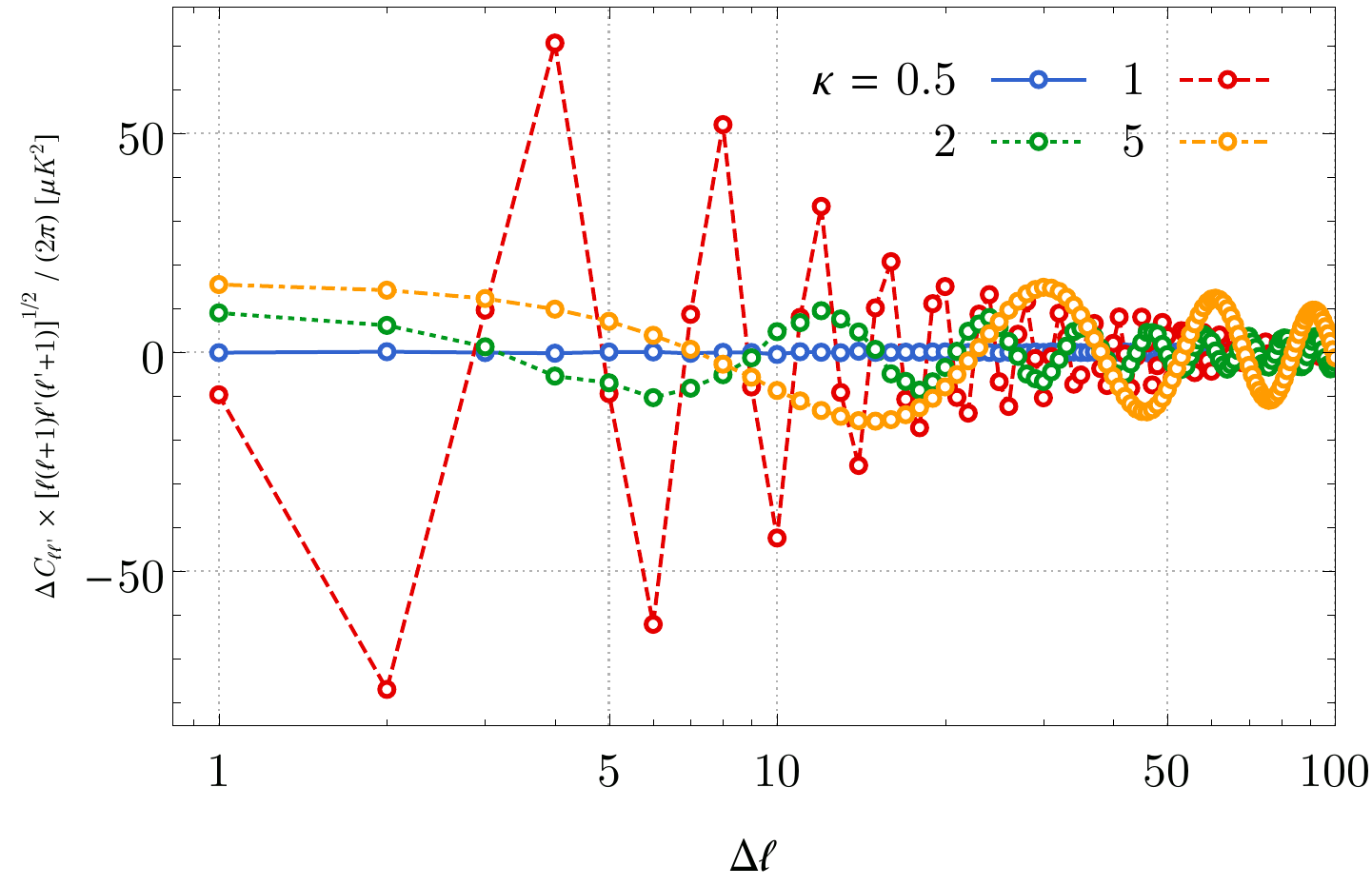}
\hspace{1cm}
\includegraphics[ width=0.46\linewidth]{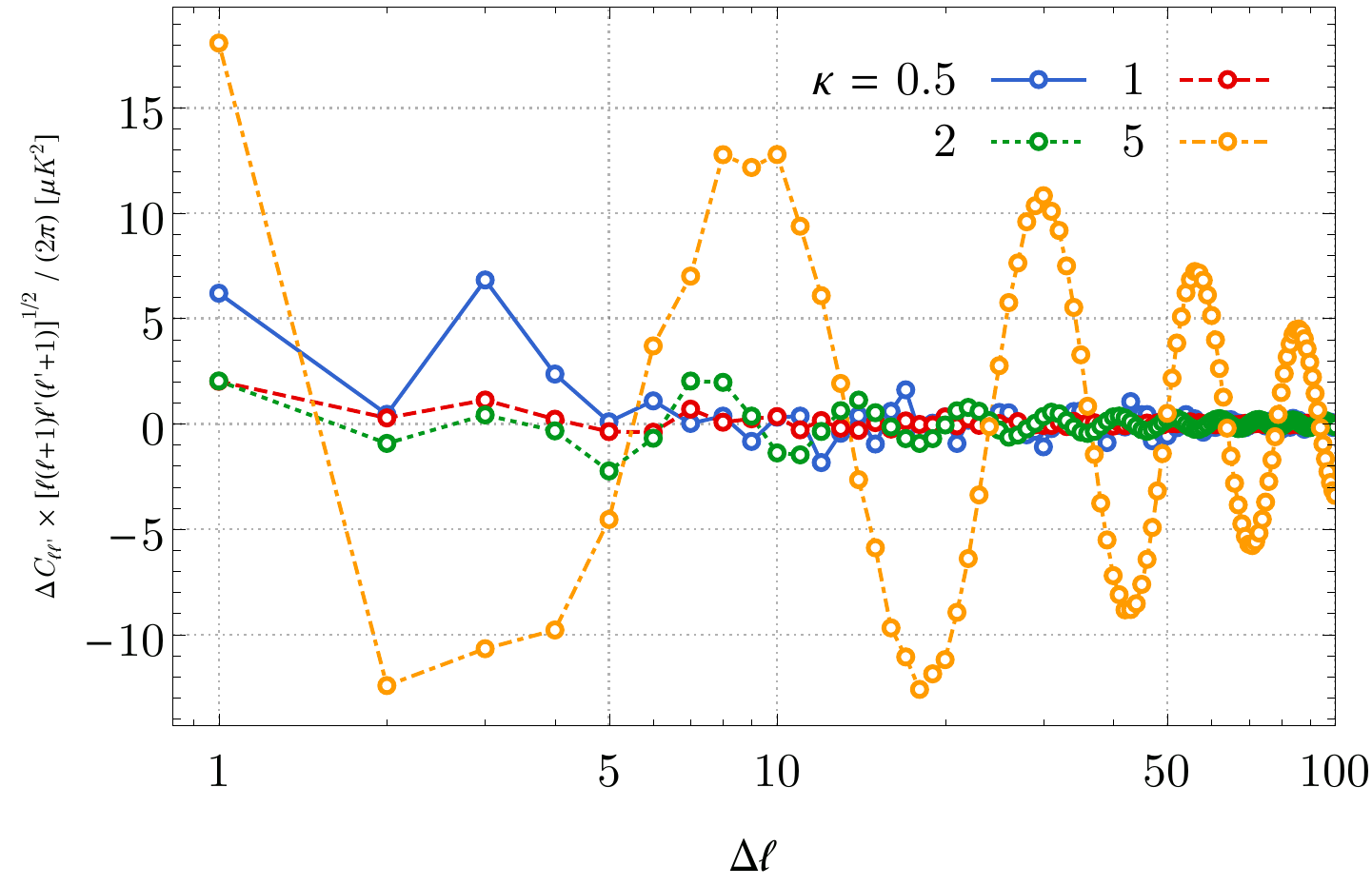}
\caption{Contributions of the domain wall to the off-diagonal elements of the CMB correlation matrix $C_{\ell\ell'}$ for two fixed values of $\ell_1$, two of $m$, and four of the parameter $\kappa$, plotted against $\Delta\ell=\ell_2-\ell_1$. {\bf Upper left:} We set $\ell_1=3$, $m=0$. {\bf Upper right:} We set $\ell_1=m=3$. {\bf Lower left:} We set $\ell_1=50$, $m=0$. {\bf Lower right:} We set $\ell_1=m=50$. In each panel the plotted quantities correspond to $\langle a_{\ell_1, m} a_{\ell_2, m}\rangle$ in Eq.~(\ref{covmat}), where we set $\ell_2=\ell_1+\Delta\ell$.}
\label{ell3ell50m0mell-fig}
\end{figure}

In Fig.~\ref{ell3ell50m0mell-fig}, we plot instead the off-diagonal elements of the covariance matrix, generated purely by the domain wall,
\begin{align}
  \Delta C_{\ell_1 \ell_2} \equiv 
-(4\pi)^2 \mathcal{P}_0 2\pi^2  \beta ~ \mathcal{T}(\ell_1, \ell_2, m, m)\label{cellellprime},
\end{align}
against $\Delta\ell\equiv\ell_2-\ell_1$. We fix $\ell_1$ in Eq.~(\ref{cellellprime}) to specific values ($3$ and $50$). In each case two plots have been made, one for $m=0$ and the other for $m=\ell_1$. We observe from all four plots that the magnitudes of the domain wall contributions to the off-diagonal power spectra,
$\Delta C_{\ell_1\ell_2}$, oscillate but decay as we increase $\Delta\ell$. This shows that the off-diagonal elements of the covariance matrix become negligible when they are far from the diagonal.

The cases plotted in Figs.~\ref{dell0m0mell-fig} and \ref{ell3ell50m0mell-fig} are instructive, demonstrating the complicated structure of the covariance matrix in our model, and present some interesting features. We should, however, emphasize here that in order to properly test the model using observational data and constrain its parameters, one will need to work with the full covariance matrix as given by Eqs. (\ref{covmat}) and (\ref{eq:calc}). The standard procedure is to construct a likelihood for the model based on this covariance matrix and scan over the full parameter space. This should contain the cosmological parameters together with the two new parameters of our domain wall model, $\beta$ and $\kappa$. Such an analysis is a natural and very interesting next step. However, it will require an extensive statistical analysis where the complications of working with real data are appropriately taken into account. Such an extensive statistical study of the model is beyond the scope of this paper and we leave it for future work. 

\section{Conclusions and outlook}
\label{conc}

In this paper we have studied an inflationary scenario in which translational invariance is broken along one direction, due to the presence of a domain wall during inflation. To simplify the analysis, we have assumed that the domain wall disappeared by the  end of inflation: it was dissolved by a currently-unspecified mechanism so that the Universe after inflation can be safely modeled by the standard isotropic cosmology. Consequently, it has been assumed that all the effects of the domain wall on cosmological observables are contained in curvature perturbations on superhorizon scales. Therefore, the picture presented here is minimal. In principle, one can consider a more realistic scenario in which the domain wall was generated dynamically during inflation from some symmetry-breaking mechanism, such as the waterfall mechanism in hybrid inflation \cite{Linde:1993cn}.  However, this makes the analysis highly complicated as one has to, for example, take into account the dynamics of the waterfall (in the waterfall scenario) and the process in which the domain wall had been created and dissolved later, presumably after reheating.

We have calculated various corrections to the power spectrum of curvature perturbations in this simple setup, assuming that the dominant source of energy is from the inflaton potential and that the contributions of the domain wall are only subleading. The domain wall breaks translational invariance, so the induced power spectrum from the wall changes along the direction perpendicular to its plane. The model is parameterized by two quantities, $\beta$ and $\kappa$. The former is a measure of the wall's tension while the latter specifies the position of the CMB sphere relative to the wall.

We studied the structure of the new contributions to the two-point correlations of the curvature perturbations in both Fourier space (as we called them, power spectra) and real space in terms of the two-point correlation function and variance. We observed that the power induced by the domain wall is scale-dependent and generates dipole, quadrupole, and higher multipoles to the power spectrum and to the variance of fluctuations on the CMB sky, and can therefore provide a mechanism to explain the anomalous power asymmetry observed by the {\it WMAP} and {\it Planck} experiments.

Our results show that for cases where the entire CMB sphere is located on one side of the domain wall ($\kappa<1$), the model gives a dipole asymmetry that is dominant over quadrupole, octopole, and higher multipoles, as expected from observations \cite{Akrami:2014eta}. For cases where the CMB sphere intersects the domain wall, the amplitudes of at least some of the higher multipoles are too large to be consistent with observational measurements. For $\kappa\gtrsim2$, either the quadrupole or octopole (or both) dominates over the dipole, and we can straightaway rule out the model for these values of $\kappa$. In addition, for $1\lesssim\kappa\lesssim2$, either the quadrupole or octopole are unacceptably large compared to the dipole amplitude, while higher multipoles (especially $a_4$ and $a_5$) oscillate and can also take on large values, so we also exclude models with these values of $\kappa$. Consequently it seems that we can safely rule out the scenario in which the domain wall intersected the CMB sphere.

Clearly, as discussed in the body of the paper, in order to test the viability of this model in comparison to the observational data and constrain its parameters in a statistically appropriate way, one needs to perform a likelihood analysis where the parameter space of the model is scanned over. Following the standard recipes, this can be done when the full covariance matrix of the model in spherical harmonic space is available. Our model is an anisotropic one and therefore the non-diagonal elements of the covariance matrix do not vanish. This means that one cannot work with only the modified angular power spectrum $C_\ell$ to perform a likelihood analysis. We have therefore derived the full covariance matrix of the model in terms of the CMB multipoles and the model parameters. We discussed some interesting features of the corrections to the power spectrum as well as to the non-diagonal elements of the covariance matrix by plotting some of these as functions of multipole. An extensive statistical study of the model requires huge computational power and is beyond the scope of this paper; we therefore leave this for future work.

In summary, we have proposed a concrete, primordial mechanism for producing an anisotropic universe which can explain the power asymmetry observed in the CMB data. This model seems to be able to naturally produce asymmetry at the level detected by CMB experiments with reasonable choices of model parameters. Although the structure of the asymmetry proposed here is too complicated to be modeled by a simple dipole modulation, which is the most widely used phenomenological explanation of the power asymmetry, our model effectively provides a dipole modulation with a scale-dependent amplitude, which is in fact a better fit to observations than the simple all-scale modulation. Further investigations of the predictions of the model for both the CMB and large-scale structure should confirm its viability or rule it out. This will be done in future work.

\acknowledgments
We would like to thank A. A. Abolhasani, Iain Brown, Hans Kristian Eriksen, Yabebal Fantaye, Frode Hansen and Mikjel Thorsrud for discussions. Y.A. is supported by the European Research Council (ERC) Starting Grant StG2010-257080. A.R.S. is supported by the David Gledhill Research Studentship, Sidney Sussex College, University of Cambridge; and by the Isaac Newton Fund and Studentships, University of Cambridge. Y.W. is supported by a Starting Grant of the European Research Council (ERC STG Grant No. 279617) and the Stephen Hawking Advanced Fellowship.

\bibliography{references}

\end{document}